\documentclass[apj]{emulateapj}
\usepackage{apjfonts}

\newcommand{\myemail}{matsuoka@kusastro.kyoto-u.ac.jp}

\shorttitle{FIR and Accretion Luminosities of the Present-Day AGNs}
\shortauthors{Matsuoka et al.}

\begin{document}



\title{Far-Infrared and Accretion Luminosities of the Present-Day Active Galactic Nuclei}






\author{Kenta Matsuoka\altaffilmark{1,2} and Jong-Hak Woo\altaffilmark{1}}



\altaffiltext{1}{Department of Physics and Astronomy, Seoul National University, 599 Gwanak-ro, Gwanak-gu, Seoul 151-742, Republic of Korea; \myemail \ and woo@astro.snu.ac.kr}
\altaffiltext{2}{Department of Astronomy, Kyoto University, Kitashirakawa-Oiwake-cho, Sakyo-ku, Kyoto 606-8502, Japan}



\begin{abstract}
We investigate the relation between star formation (SF) and black hole accretion luminosities, using a sample of 492 type-2 active galactic nuclei (AGNs) at $z < 0.22$, which are detected in the far-infrared (FIR) surveys with AKARI and {\it Herschel}. We adopt FIR luminosities at 90 and 100 \micron\ as SF luminosities, assuming the proposed linear proportionality of star formation rate with FIR luminosities. By estimating AGN luminosities from [\ion{O}{3}]$\lambda$5007 and [\ion{O}{1}]$\lambda$6300 emission lines, we find a positive linear trend between FIR and AGN luminosities over a wide dynamical range. This result appears to be inconsistent with the recent reports that low-luminosity AGNs show essentially no correlation between FIR and X-ray luminosities, while the discrepancy is likely due to the Malmquist and sample selection biases. By analyzing the spectral energy distribution, we find that pure-AGN candidates, of which FIR radiation is thought to be AGN-dominated, show significantly low-SF activities. These AGNs hosted by low-SF galaxies are rare in our sample ($\sim 1\%$). However, the low fraction of low-SF AGN is possibly due to observational limitations since the recent FIR surveys are insufficient to examine the population of high-luminosity AGNs hosted by low-SF galaxies.
\end{abstract}






\section{Introduction}\label{int}
In decades the connection between galaxy evolution and the growth of supermassive black holes (SMBHs) has been one of the main topics in extragalactic research. The tight correlation of black hole mass, $M_{\rm BH}$, with galaxy properties, e.g., stellar velocity dispersion, $\sigma_{*}$ \citep[e.g.,][]{1998AJ....115.2285M,2003ApJ...589L..21M,2010ApJ...716..269W,2013ApJ...772...49W}, suggests the coevolution of galaxies and SMBHs although the physical link between them is yet to be clearly revealed. Various observational studies have been devoted to investigating the nature of the coevolution. For example, the redshift evolution of the $M_{\rm BH}$-$\sigma_{*}$ relation, representing a cumulative growth history, has been investigated mainly using type-1 active galactic nuclei, AGNs \citep[e.g.,][]{2006ApJ...645..900W,2008ApJ...681..925W,2010ApJ...708..137M,2013ApJ...767...13S}. The connection between on-going star formation (SF) and AGN activity is also one of the observational signatures, revealing the connection of the growth of stellar mass and the BH growth at the observed epoch \citep[e.g.,][]{1988ApJ...325...74S,2001ApJ...558...81C,2003MNRAS.346.1055K,2009MNRAS.399.1907N,2012NewAR..56...93A,2012A&A...545A..45R}.

Various theoretical frameworks have been suggested to explain the AGN-SF link and also reproduce the $M_{\rm BH}$-$\sigma_{*}$ relation. For example, based on the smoothed particle hydrodynamic $N$-body simulations of gaseous galaxy mergers, \citet{2011MNRAS.412.2154B} showed simultaneous bursts of SF and BH accretion \citep[see also][]{2010MNRAS.407.1529H}. Such theoretical models predict a positive correlation between SF and AGN activity, as consistent with the results of several observational studies \citep[e.g.,][]{2007ApJ...666..806N,2009MNRAS.399.1907N,2012ApJ...746..168D,2012A&A...546A..58R,2014ApJ...784..137K}. Note that other studies reported that there is a time lag between SF and AGN phases \citep[e.g.,][]{2007ApJ...671.1388D,2010MNRAS.405..933W,2011A&A...527A.100M,2012ApJ...755...28C,2012MNRAS.420L...8H}.

While the AGN-SF link can be investigated in various aspects, the direct comparison between AGN and SF luminosities, i.e., the $L_{\rm AGN}$-$L_{\rm SF}$ relation, is the most simple approach. Since the SF luminosity corresponds to on-going growth of galaxies and the AGN luminosity reveals the current growth of SMBHs, the AGN-SF connection can be directly traced at the observed epoch. A correlation between SF and AGN luminosities has been reported in the previous studies, indicating that luminous AGNs are hosted by highly star-forming galaxies. Based on the combined sample of local type-2 AGNs and quasars at $0.1 \leq z < 3$, for example, \citet{2009MNRAS.399.1907N} found there is a good correlation between SF and AGN luminosities, albeit with substantial scatters \citep[see also][]{2007ApJ...666..806N,2008ApJ...684..853L,2012AJ....143...49W}. Recently, \citet{2012ApJ...753..155T} have found a correlation between $L_{\rm AGN}$ and $L_{\rm IR}$ of low-ionization nuclear emission-line regions (LINERs). These results suggest that BH activity is connected with SF.

Based on the deep {\it Herschel} imaging of the X-ray sources at $0.2 < z < 2.5$ in the fields of GOODS and COSMOS, \citet{2012A&A...545A..45R} have reported a correlation between AGN and FIR luminosities. In their study, luminous X-ray AGNs at $z < 1$ show a correlation between AGN and FIR (i.e., 60 \micron) luminosities, as similarly presented by earlier works \citep[e.g., ][]{2007ApJ...666..806N,2008ApJ...684..853L,2009MNRAS.399.1907N} while the correlation flattens or disappears at $z > 1$ \citep[see also, e.g.,][]{2010ApJ...712.1287L,2010A&A...518L..26S,2010A&A...518L..33H,2012ApJ...760L..15H,2012Natur.485..213P}. In contrast, they claimed that low-luminosity AGNs show essentially no correlation between FIR and AGN luminosities at all redshift. The enhanced SF for given AGN luminosity of their low-$z$ X-ray AGNs ($0.2 < z < 0.5$) seems to contrast to the finding of \citet{2009MNRAS.399.1907N} that local type-2 AGNs ($z \leq 0.2$) show a positive correlation between SF and AGN luminosities. This discrepancy may be caused by observational biases, e.g., the Malmquist and sample selection biases, and from the measurement uncertainties in SF and AGN luminosities. It is also possible that for a given AGN luminosity, galaxies with lower SF luminosity may be undetected due to the FIR flux limit. Therefore, in order to fully reveal the connection between AGN and SF activities it is crucial to examine potential biases, which may affect the $L_{\rm FIR}$-$L_{\rm AGN}$ relation.

FIR luminosity is often used as a SF indicator since the rest-frame FIR emission is mainly from the host galaxy while the AGN contribution to FIR, $\sim$ 50$-$150 \micron, is due to the Rayleigh-Jeans tail of an AGN-heated dust component \citep[e.g.,][]{2009MNRAS.399.1907N,2012MNRAS.419...95M,2012A&A...545A..45R}. In the case of AGN luminosity, X-ray luminosity can be used with a proper bolometric correction. However, X-ray data with sufficient depth is often not available. Instead, emission lines from the narrow-line regions (NLRs), e.g., H$\beta$, [\ion{O}{3}]$\lambda$5007, [\ion{O}{1}]$\lambda$6300, and [\ion{O}{4}]$\lambda$25.89\micron\ lines, are often used as a proxy for AGN bolometric luminosity \citep[e.g.,][]{2009MNRAS.399.1907N,2012ApJ...746..168D}.

In this paper, we investigate the AGN-SF connection for a sample of type-2 AGNs at $0.01 \leq z < 0.22$ selected from the Sloan Digital Sky Survey (SDSS), using AKARI and {\it Herschel} data. In Section~\ref{sam}, we describe the sample selection and the data. Section~\ref{res} presents the main results and Section~\ref{dis} provides discussion and interpretation. The summary and conclusion are given in Section~\ref{con}. We adopted a concordance cosmology with ($\Omega_{\rm M}$,$\Omega_{\rm \Lambda}$) = (0.3,0.7) and $H_{\rm 0}$ = 70 km s$^{-1}$ Mpc$^{-1}$.

\section{Sample and data}\label{sam}
In this study, we mainly focus on the local type-2 AGNs selected from SDSS, for which FIR data are available. Using SDSS spectroscopic data and FIR survey data, we investigate the relation between AGN and FIR luminosities in the wide dynamic range. We also collect multi-wavelength data, e.g., ultraviolet (UV) and mid-infrared (MIR), to examine the characteristics of the galaxies in the sample. In this section, we describe our sample selection and multi-wavelength data.

\subsection{Type-2 AGN Sample}\label{type}
We selected type-2 AGNs at $0.01 \leq z < 0.22$ from the MPA-JHU SDSS DR7 galaxy catalog\footnote{http://www.mpa-garching.mpg.de/SDSS/}, including $927,552$ galaxies, based on the BPT diagnostic diagram \citep[e.g.,][]{1981PASP...93....5B} using the [\ion{O}{3}]$\lambda$5007/H$\beta$ and [\ion{N}{2}]$\lambda$6584/H$\alpha$ flux ratios with the classification scheme in \citet{2006MNRAS.372..961K}. Note that we excluded composite objects since they are unreliable in estimating AGN luminosities from narrow-emission lines due to the contribution from SF. In the selection process, we also adopted following criteria: reliable redshift measurement (i.e., $z_{\rm warning} = 0$) and high signal-to-noise ratios (S/N) for emission lines which were used in AGN classification, i.e., S/N $> 6.0$ for [\ion{O}{3}]$\lambda$5007, H$\alpha$, and [\ion{N}{2}]$\lambda$6584 lines, S/N $> 3.0$ for the H$\beta$ line. As a result, we obtained 35,945 type-2 AGNs.

Note that our sample contains both Seyfert 2s and LINER 2s, although the physical connection between them is not clear as discussed in various studies. Especially understanding the ionization process of LINERs is essential in our study because it directly relates to the estimate of AGN luminosities (Section~\ref{arel}). Mainly there are three considerations for LINER 2s: (1) hot stars, i.e., post-AGB star and blue stars showing similar line-flux ratios to LINERs, (2) shock ionizations instead of photoionozations, and (3) low-ionization parameters of LINER 2s. In order to consider these issues, we divide our sample into two subsamples, i.e., Seyfert 2s and LINER 2s (Section~\ref{fari}). To remove stars showing similar line-flux ratios to LINERs, we identify the so-called [\ion{O}{1}]$\lambda$6300-weak LINERs, which are considered non-AGNs. By adopt a criterion, i.e., [\ion{O}{1}]$\lambda$6300/H$\alpha$ $< 1/6$ \citep[e.g.,][]{1992ApJ...397L..79F,2000ApJ...530..688A}, we exclude LINER 2s (see Section~\ref{fari}). Regarding shock ionizations, we expect that shocked gas should lead to higher excitation UV spectra than photoionized gas. Thus, strong UV emission lines, e.g., \ion{C}{4}$\lambda$1549, are expected. However, some observational studies reported featureless UV spectra of LINERs. For example, \citet{1996AJ....112.1829B,1997AJ....114.2313B} presented UV spectra of local LINERs that high-excitation lines are not detected, meaning the fast shock models are poor matched to the observed spectra \citep[see also][]{1998AJ....116...55M,1998MNRAS.300..893N,2000ApJ...532..883G,2003ApJ...584..164S}. In this study, therefore we assume that the majority of our LINERs are photoionized objects. Moreover, the low-ionization parameter of LINERs is another concern to consider. Since the ionization parameters of LINERs are systematically smaller than that of Seyferts, there is large uncertainties in determining AGN luminosity adopting the same method used for Seyferts. However, if we use the method based on both [\ion{O}{3}]$\lambda$5007 and [\ion{O}{1}]$\lambda$6300 lines, we can correct for the ionization effect of low-ionization sources such as LINERs (see Section~\ref{arel}).

\begin{figure}
\epsscale{1.0}
\plotone{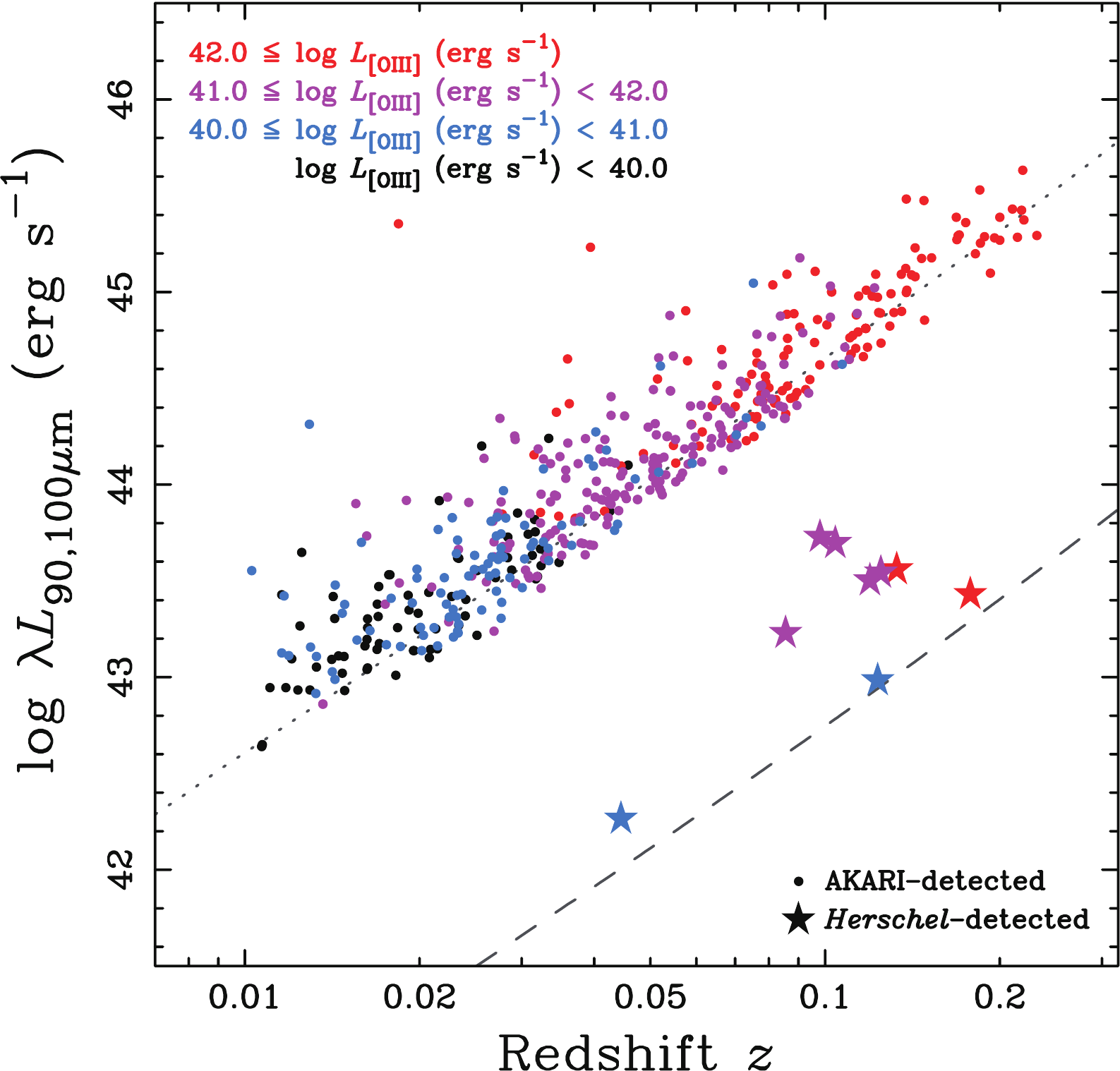}
\caption{FIR luminosity of the AKARI-detected (90 \micron; small circles) and {\it Herschel}-detected sources (100 \micron; stars) as a function of redshift. The sample is color-coded based on the luminosity of [\ion{O}{3}]$\lambda$5007 line, as labeled at the top left. The dotted and dashed lines represent the 5$\sigma$-detection limits of the AKARI/FIS and the {\it Herschel}/PACS surveys, respectively.}
\label{adam}
\end{figure}

\subsection{FIR Data from AKARI and Herschel}\label{fari}
To obtain FIR luminosities, we first cross-identified AGNs against the AKARI/FIS all-sky survey bright source catalog \citep{2010yCat.2298....0Y}. AKARI is the first Japanese infrared astronomical satellite \citep{2007PASJ...59S.369M} which carries two instruments, i.e., the Infrared Camera \citep[IRC;][]{2007PASJ...59S.401O} and the Far-Infrared Surveyor \citep[FIS;][]{2007PASJ...59S.389K}.

The all-sky survey has been performed with FIS in four bands, respectively, centered at 65, 90, 140, and 160 \micron. In this study, we utilised 90 \micron\ sources with the flux quality flag of FQUAL $= 3$ (i.e., high quality). The 5$\sigma$-detection limit of the 90 \micron\ band is 0.55 Jy. By matching the SDSS AGNs with the 90 \micron\ sources within the maximum radius of 18\arcsec, which corresponds roughly to the 3$\sigma$ position error in the cross-scan direction of FIS, we obtained 678 AKARI/FIS counter parts of the type-2 AGNs, which is $\sim$ 1.9\% of all AGNs in the sample. Note that we checked SDSS spectra of whole FIR-detected AGNs based on visual inspection in order to avoid unusable data, e.g., noisy data.

To overcome the shallow flux limit of the FIS survey, we additionally used the PACS Evolutionary Probe (PEP) survey data \citep{2011A&A...532A..90L}. PEP is a deep FIR photometric survey with the Photodetector Array Camera and Spectrometer \citep[PACS;][]{2010A&A...518L...2P}, on board of the {\it Herschel} Space Observatory \citep{2010A&A...518L...1P}. The field selection of the PEP survey includes popular multi-wavelength fields such as GOODS, COSMOS, Lockman Hole, ECDFS, and EGS. Twelve objects in our AGN sample are located in the COSMOS field \citep{2007ApJS..172....1S}, and none of them were detected in the AKARI survey. By matching these 12 objects with the PEP 100 \micron\ source catalog, which has 5$\sigma$-detection limit 0.0075 Jy, we obtained FIR counterparts for 11 objects.

As described in Section~\ref{type} we divide our FIR sample into two subsamples of Seyfert 2s and LINER 2s, and remove [\ion{O}{1}]$\lambda$6300-weak LINERs. First, we adopted criteria based on the BPT diagram using the [\ion{O}{3}]$\lambda$5007/H$\beta$ and [\ion{O}{1}]$\lambda$6300/H$\alpha$ flux ratios \citep[i.e.,][]{2006MNRAS.372..961K} to separate the sample into 355 Seyfert 2s and 231 LINER 2s, including six and three {\it Herschel}-detected objects, respectively. In this classification, we gave a criterion, S/N $> 3.0$ for the [\ion{O}{1}]$\lambda$6300 line. Then, adopting a criterion, i.e., [\ion{O}{1}]$\lambda$6300/H$\alpha$ $< 1/6$ \citep[e.g.,][]{1992ApJ...397L..79F,2000ApJ...530..688A} we removed 94 [\ion{O}{1}]$\lambda$6300-weak LINERs.

In summary, we obtained 483 AKARI/FIS-detected objects and nine {\it Herschel}/PACS-detected objects, for which we investigate the AGN-SF connection in next sections. Figure~\ref{adam} presents the FIR-luminosity distribution of our sample as a function of redshift, clearly showing that PEP survey is almost two orders of magnitude deeper and complementary to the shallow FIS survey sample.

\subsection{MIR Data from WISE}\label{midi} 
We collected MIR data of 483 AKARI-detected objects, by matching them against the All{\it WISE} source catalog\footnote{http://wise2.ipac.caltech.edu/docs/release/allwise/}, which is the most recent data release of the Wide-field Infrared Survey Explorer \citep[{\it WISE};][]{2010AJ....140.1868W}, after combing all previous data from the {\it WISE} cryogenic and NEO{\it WISE} \citep{2011ApJ...731...53M}. We adopted the maximum radius of 6\farcs1, 6\farcs4, 6\farcs5, and 12\farcs0, respectively for 3.4, 4.6, 12, and 22 \micron\ band images, accounting for the averaged point-spread functions (PSFs) in each band. For undetected sources, upper limits from the 5$\sigma$ sensitivity are given as 0.08, 0.11, 1, and 6 mJy respectively for 3.4, 4.6, 12, and 22 \micron\ bands. In this process, we obtained the MIR fluxes for 99\% of the AKARI-detected AGNs.

\subsection{UV Data from GALEX}\label{ultr} 
We also collected near-ultraviolet (NUV) and far-ultraviolet (FUV) flux data obtained by the Galaxy Evolution Explorer All-Sky-Imaging Survey \citep[{\it GALEX}/AIS;][]{2005ApJ...619L...1M}. Using the {\it GALEX}/AIS-SDSS matched catalogs in \citet{2011MNRAS.411.2770B}, which was constructed by matching the {\it GALEX} GR5 data against the SDSS DR7 with a radius of 3\arcsec, we obtained 242 and 149 counterparts out of 483 AGNs, respectively, in the NUV and FUV. For the remaining objects, an upper limit is given based on typical depths of 20.8 and 19.9 $AB$ magnitude in the NUV and FUV.

\section{Results}\label{res}
In this section, first, we compare four different SF indicators, namely, FIR luminosity, the break at 4000\AA\ (D4000), UV-luminosity, and the [\ion{O}{2}]$\lambda$3727 emission line luminosity, to investigate the reliability of the FIR luminosity as a SF indicator (Section~\ref{star}). Second, we investigate the relation between AGN and SF luminosities of the FIR-matched type-2 AGN sample (Section~\ref{arel}).

\begin{figure*}
\epsscale{0.32}
\plotone{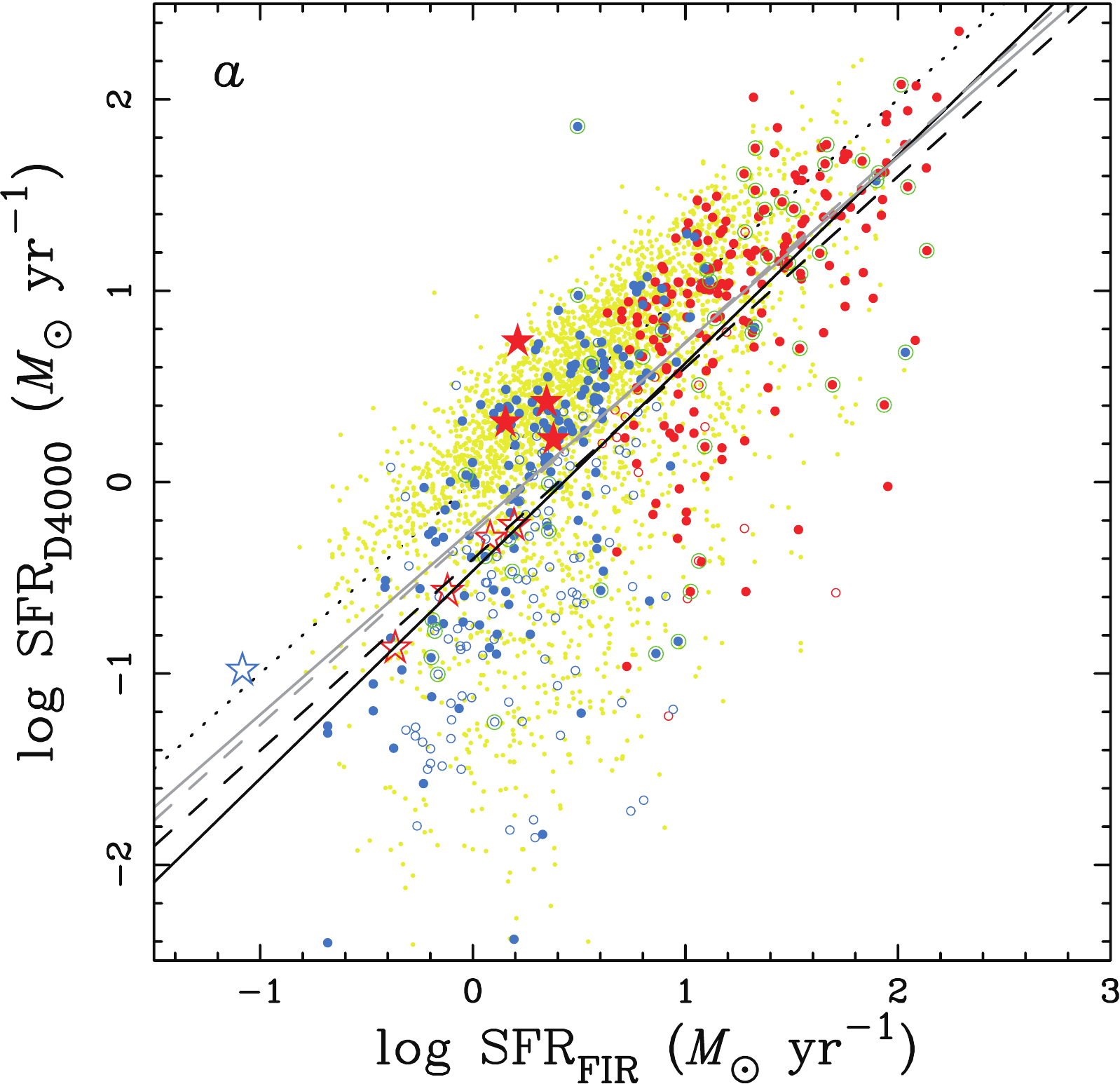}
\hspace{0.5mm}
\plotone{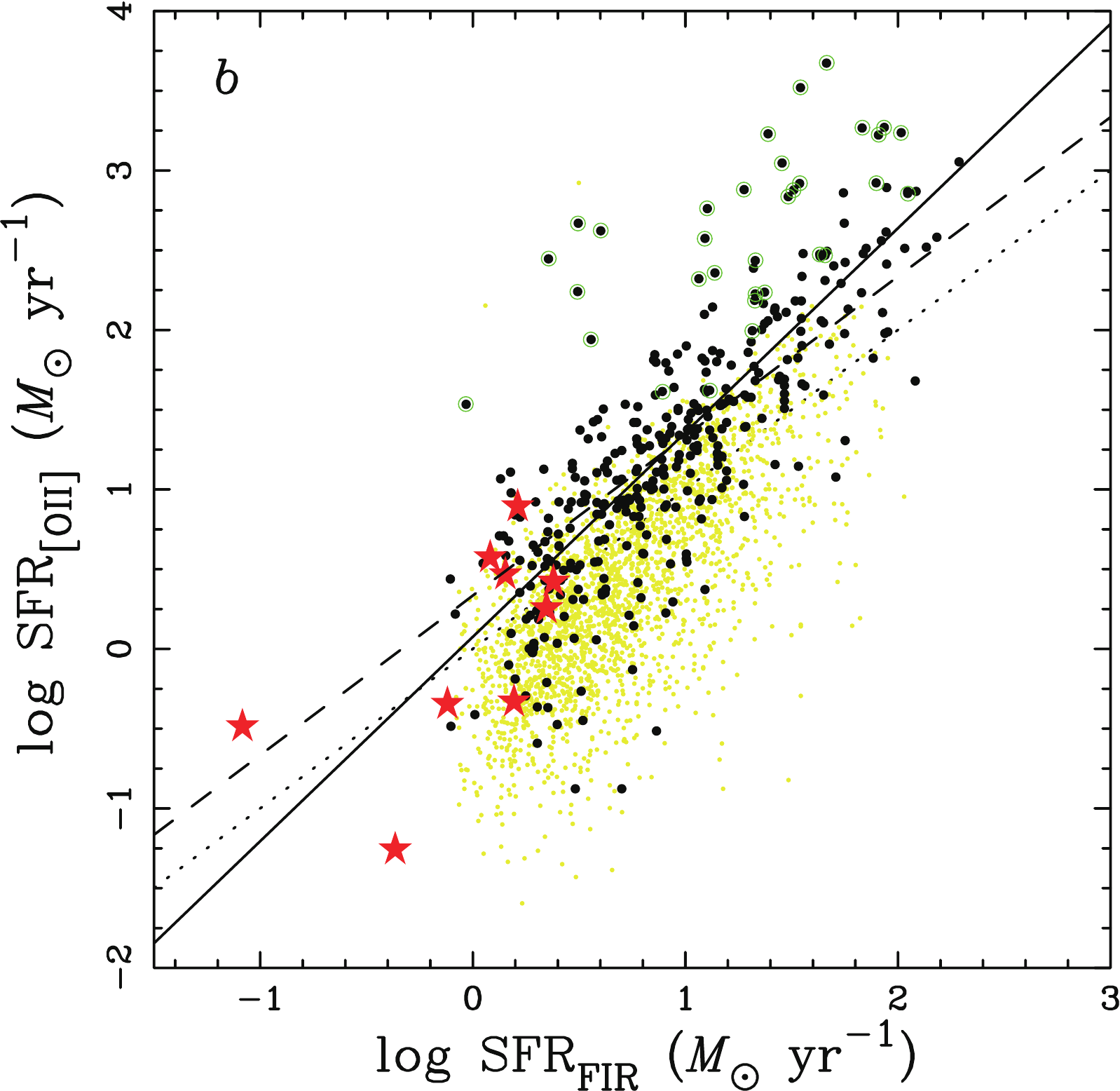}
\hspace{0.5mm}
\plotone{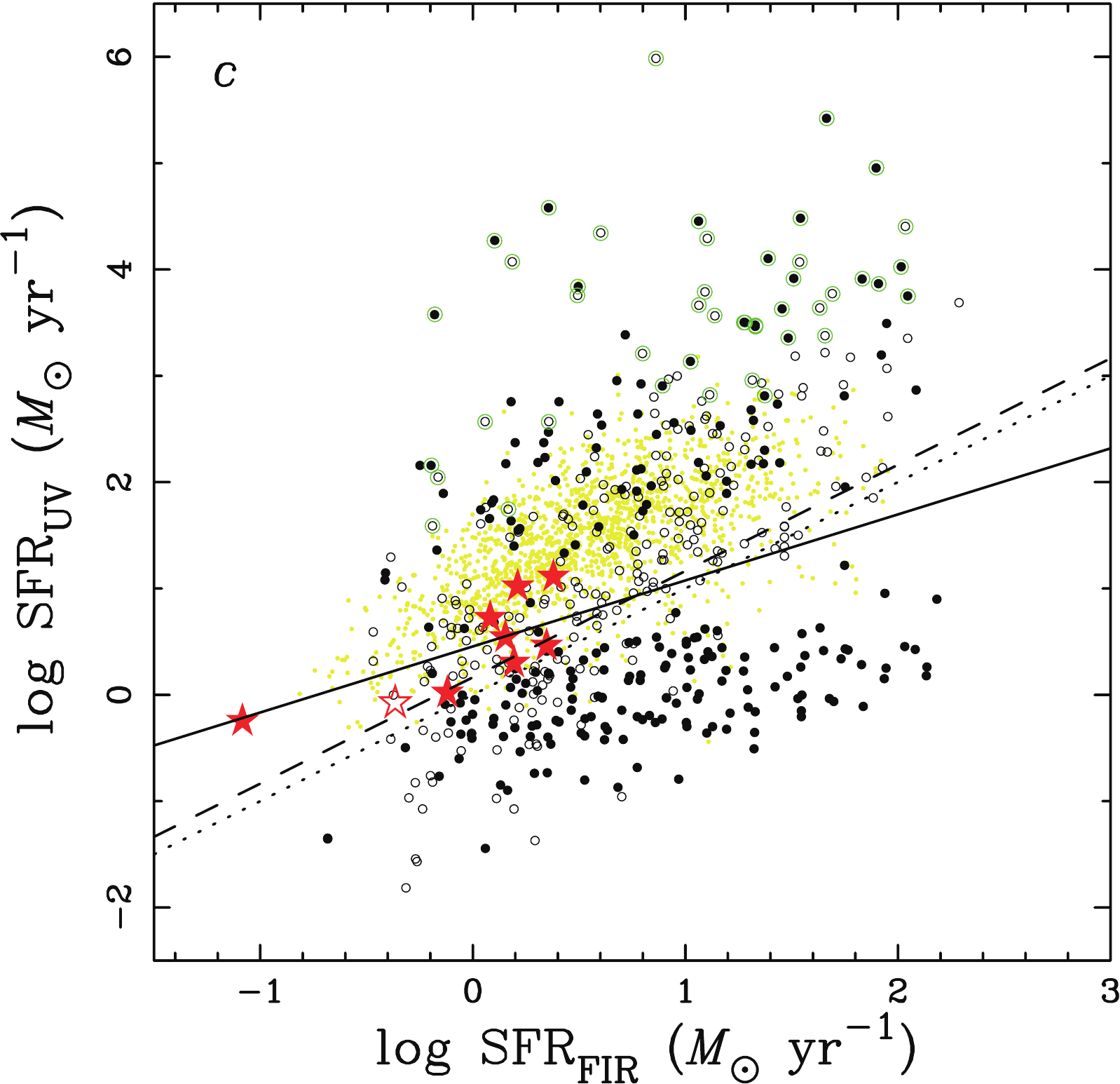}
\vspace{3.0mm}
\plotone{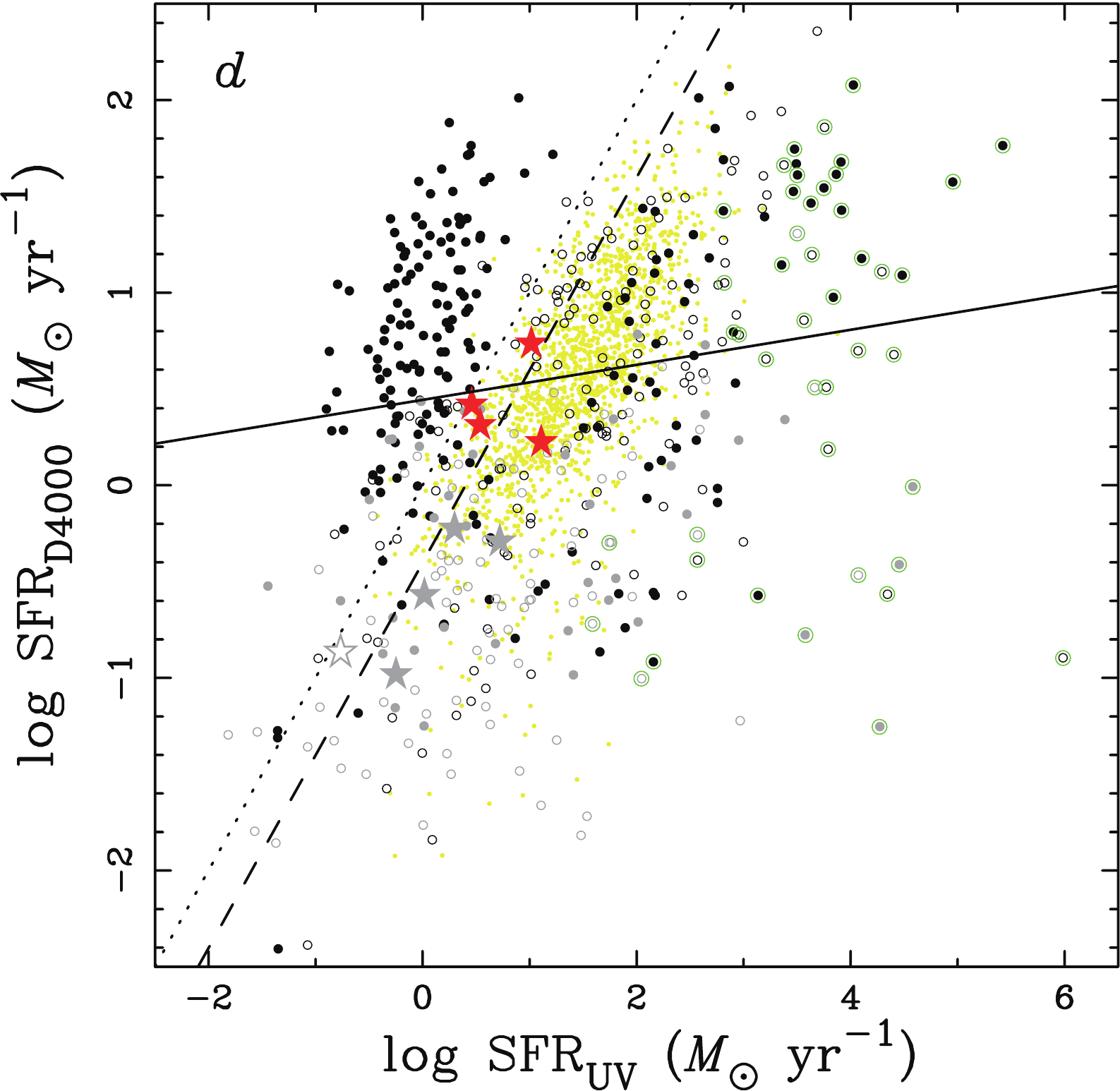}
\hspace{0.5mm}
\plotone{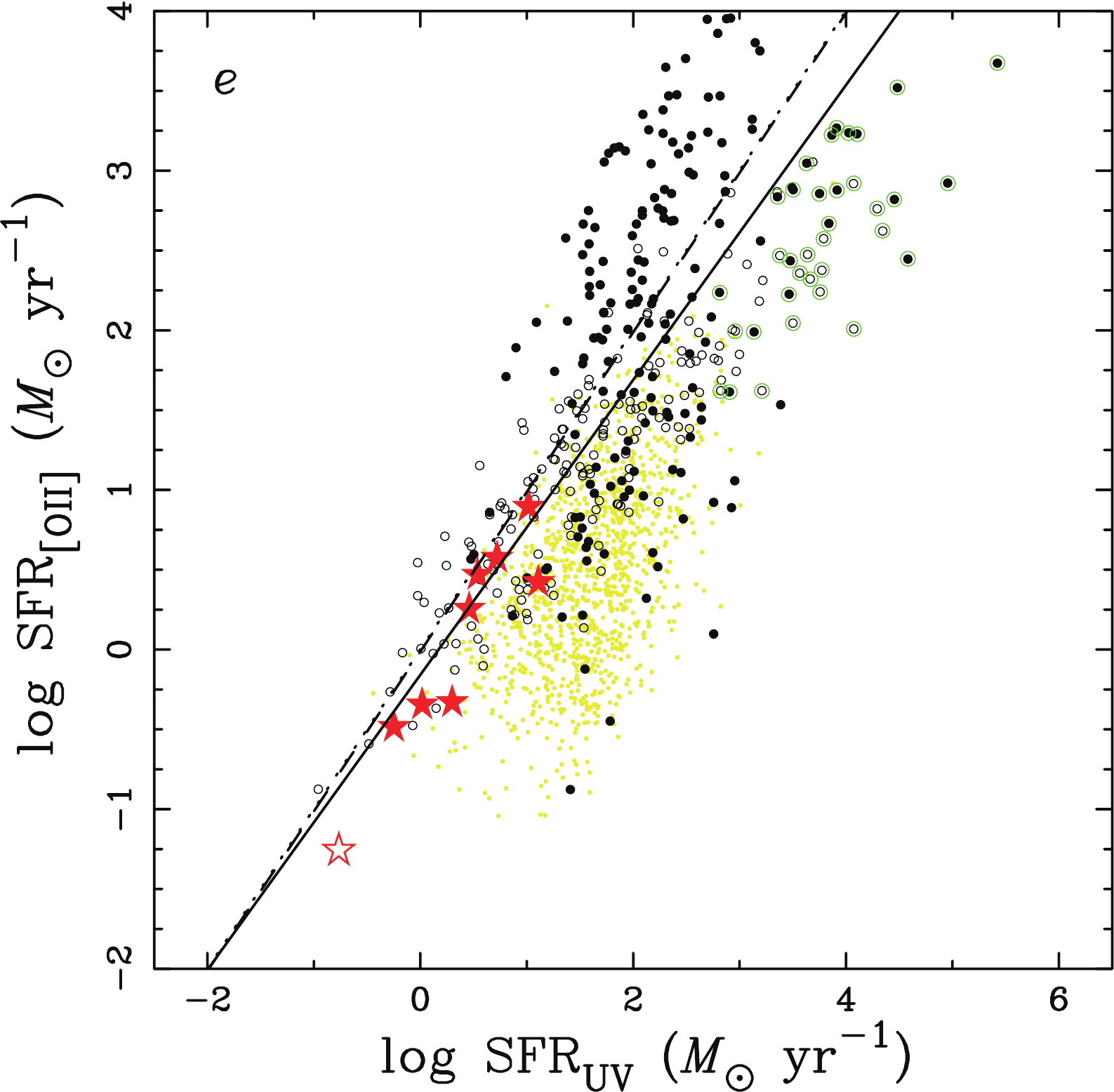}
\hspace{0.5mm}
\plotone{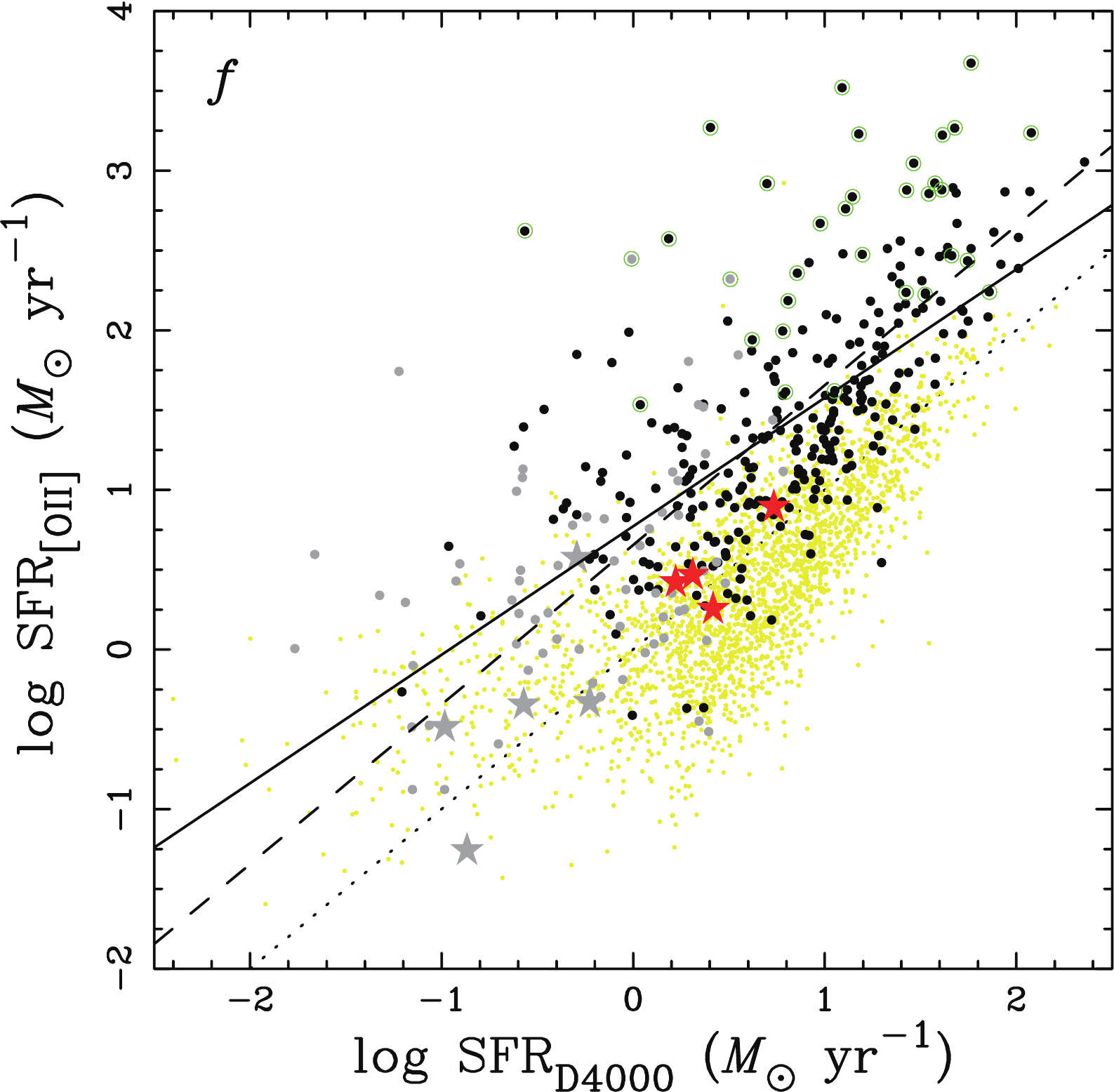}
\caption{Comparisons of the four independently estimated SFRs from FIR luminosity, D4000, UV-luminosity, and [\ion{O}{2}]$\lambda$3727 luminosity for the AKARI-detected (small filled and open circles) and {\it Herschel}-detected objects (stars). In panel~(a), blue and red symbols represent low-$z$ ($0.05 < z$) and high-$z$ ($0.05 \ge z$) objects, respectively, and open symbols indicate red or old objects, i.e., D4000 $>$ 1.8. In panels~(c), (d), and (e), upper limits of the UV-based SFR are shown as open symbols. Gray symbols in panels~(d) and (f) represent the red or old population same as open symbols in panel~(a). Green-large circles mark highly obscured objects, i.e., $E_{B-V} > 1$. Gray-dotted line indicates the one-to-one relation, and black solid and dashed lines are fitting results, i.e., $y = a x + b$ and $y = x + c$, respectively. Gray solid and dashed lines in panel~(a) are best-fit for data excluding red or old objects. Yellow circles represent normal star-forming galaxies detected with AKARI.}
\label{fred}
\end{figure*}

\subsection{SF Indicators}\label{star}
To test whether the FIR luminosity $L_{\rm FIR}$ is a reasonable SF indicator, we compare $L_{\rm FIR}$ to other SF indicators, i.e., D4000 \citep{2004MNRAS.351.1151B}, UV luminosities, and [\ion{O}{2}]$\lambda$3727 line luminosity. We obtained these measurements and then converted them to star formation rates (SFRs) as explained below.

For FIR luminosity, we collected 90 and 100 \micron\ data respectively from the AKARI and {\it Herschel} samples, as described in Section~\ref{fari}. For given spectral energy distributions (SEDs) of typical SF galaxies \citep[e.g.,][]{2002ApJ...576..159D}, the difference between fluxes at 90 and 100 \micron\ is relatively small, e.g., $\log (90 F_{\rm 90\mu m}) / \log (100 F_{\rm 100\mu m}) \sim 1.005$. Thus, we used 90 and 100 \micron\ fluxes for calculating $L_{\rm FIR}$. We also ignore the redshift effects since the redshift range of our sample is small ($0.01 \le z < 0.22$): even for the highest redshift objects at $z = 0.22$ in our sample, the flux correction is negligible, i.e., $\log (90 F_{\rm 90\mu m,obs}) / \log (90 F_{\rm 90\mu m,rest}) \sim 0.987$. Here, we adopt the conversion recipe in \citet{1998ARA&A..36..189K}:
\begin{equation}
{\rm SFR}_{\rm FIR} \ (M_\odot \ {\rm year}^{-1}) = 4.5 \times 10^{-44} L_{\rm FIR} \ ({\rm erg} \ {\rm s}^{-1}).
\end{equation}

Second, we obtained the SFR determined from the break at 4000\AA\ SFR$_{\rm D4000}$ from the MPA-JHU SDSS DR7 galaxy catalog, which is based on the technique discussed by \citet{2004MNRAS.351.1151B}. First, they constructed the relation between the specific SFR measured from the H$\alpha$ line and D4000 using a sample of star-forming galaxies. Adopting this relation along with stellar masses estimated from the mass-to-light ratios, they derived SFR from D4000 for galaxies, of which emission lines are not reliable as SF indicators due to the contamination of AGN. Note that aperture corrections have been applied \citep{2004MNRAS.351.1151B}, using the resolved color information available for each galaxy. Since SFR$_{\rm D4000}$ can be determined for AGN host galaxies, it has been adopted in various studies \citep[e.g.,][]{2009MNRAS.399.1907N}.

Third, we used the UV luminosity as a SF indicator. From the {\it GALEX} UV luminosities, we calculated SFR$_{\rm UV}$ using the recipe given by \citet{1998ARA&A..36..189K}:
\begin{equation}
{\rm SFR}_{\rm UV} \ (M_\odot \ {\rm year}^{-1}) = 1.4 \times 10^{-28} L_{\rm UV} \ ({\rm erg} \ {\rm s}^{-1} \ {\rm Hz}^{-1}),
\end{equation}
where $L_{\rm UV}$ is the luminosity density integrated over the spectral range 1500$-$2800\AA. In this section, we focus on NUV data because all FUV-detected objects are detected with NUV, and converted NUV luminosities to SFR$_{\rm UV}$. We corrected for the UV extinction using the Balmer decrement\footnote{http://ned.ipac.caltech.edu/level5/Sept01/Rosa/Rosa\_appendix.html}.

Last, we adopted the [\ion{O}{2}]$\lambda$3727 line luminosity as a SF indicator using the following equation \citep{1998ARA&A..36..189K}:
\begin{equation}
{\rm SFR}_{\rm [O\!~II]} \ (M_\odot \ {\rm year}^{-1}) = 1.4 \times 10^{-41} L_{\rm [O\!~II]} \ ({\rm erg} \ {\rm s}^{-1}),
\end{equation}
where $L_{\rm [O\!~II]}$ is the luminosity of the [\ion{O}{2}]$\lambda$3727 line. We corrected for the dust extinction (see footnote~5). Note that here we focus on S/N $> 6$ objects for the [\ion{O}{2}]$\lambda$3727 line flux to increase reliability.

We compare the four independently estimated SFRs in Figure~\ref{fred}, and discuss the details as follows. First, in comparing $L_{\rm FIR}$-based and D4000-based SFRs (Figure~\ref{fred}(a)), we find a relatively good trend at high SFR regime, while many objects are below the one-to-one relation (dotted line), indicating that D4000 underestimates SFR compared FIR luminosity. Particularly at $\log {\rm SFR}_{\rm FIR} < 0$, the discrepancy becomes unacceptably large. We fit the data with a linear function, i.e., $y = a x + b$ (black-solid line) or with a fixed slope, i.e., $y = x + c$ (black-dashed line) as listed in Table~1. Since the SFR$_{\rm D4000}$ is calibrated using starburst galaxies, the D4000 method is subject to large uncertainties for older and redder galaxies \citep[see also][]{2009MNRAS.399.1907N}. In fact, such galaxies (i.e., D4000 $> 1.8$, open symbols in Figure~\ref{fred}) are mostly located below the one-to-one line. Thus, we performed a liner fit after excluding such galaxies (gray-solid and gray-dashed lines), which slightly improves the relation. Although aperture correction was adopted in \citet{2004MNRAS.351.1151B}, we further consider the aperture effect that the fixed 3\arcsec\ fiber size of the SDSS spectroscopy covers a smaller physical area of the lower-$z$ galaxies, hence, the SFR$_{\rm D4000}$ may be more underestimated than higher-$z$ galaxies. To test the aperture effect, we divided the sample into two redshift bins, i.e., $z \ge 0.05$ (red) and $z < 0.05$ (blue) in Figure~\ref{fred}(a). However, we find no significant difference between them, concluding that the aperture effect is not the origin of the discrepancy between SFR$_{\rm FIR}$ and SFR$_{\rm D4000}$. Moreover, to examine an extinction effect, we marked highly obscured objects, i.e., $E_{B-V} > 1$ (green-large circles), but we found no significant extinction-related bias. Note that at low FIR luminosities ($\sim 10^{43}$), the FIR is believed to come from diffuse dust cirrus warmed by the background starlight, not necessarily recently formed stars. In this case, FIR luminosity overestimates SFR although such low-$L_{\rm FIR}$ objects are negligible in our samples (see Figure~\ref{adam}). To demonstrate the correlation between D4000 and $L_{\rm FIR}$, we also plot normal star-forming galaxies detected with AKARI (yellow circles). In conclusion, we suggest that SFR$_{\rm D4000}$ indicator seems to have various issues, especially for older and redder galaxy populations.

\begin{figure*}
\epsscale{0.49}
\plotone{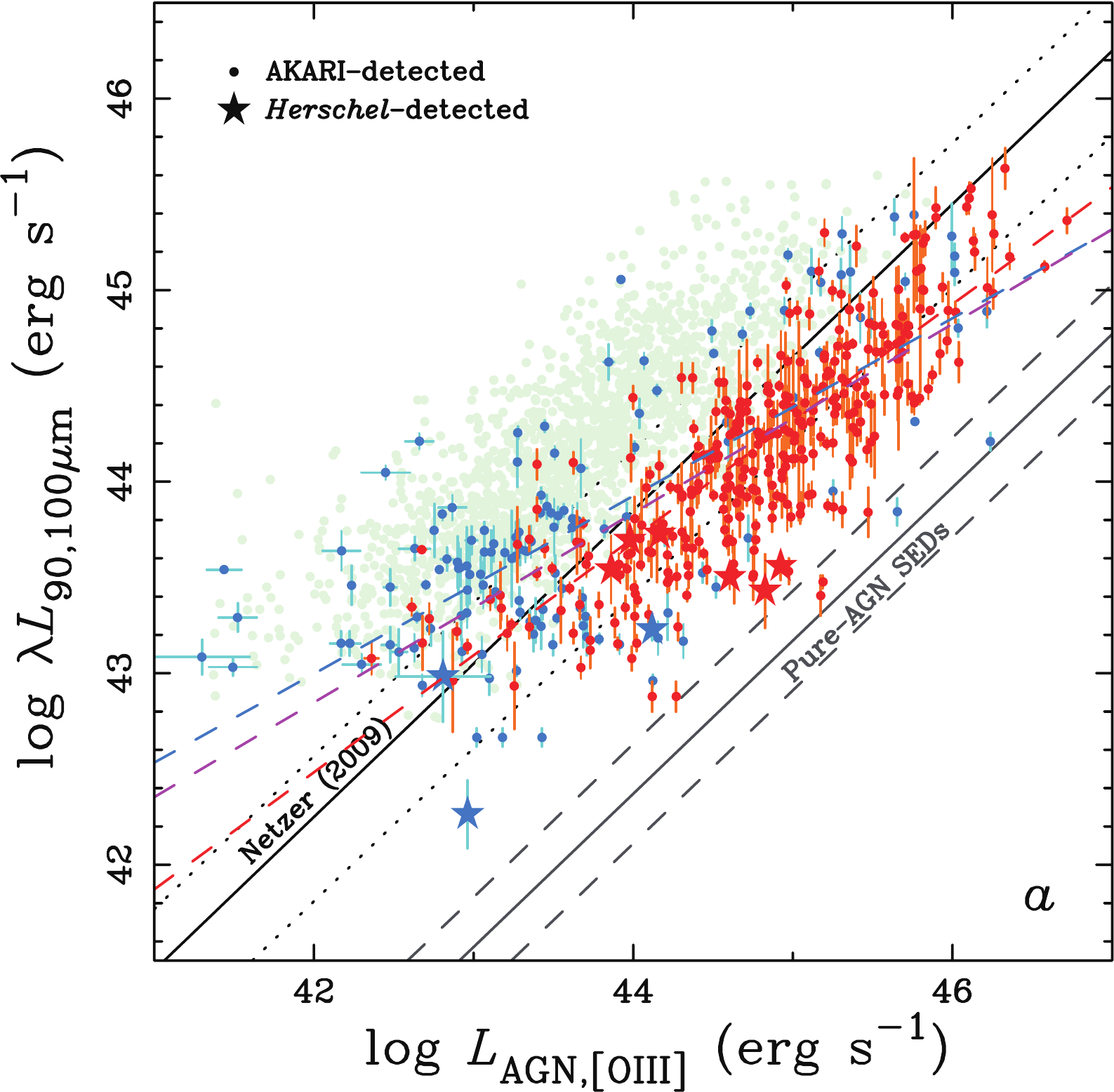}
\hspace{5.0mm}
\plotone{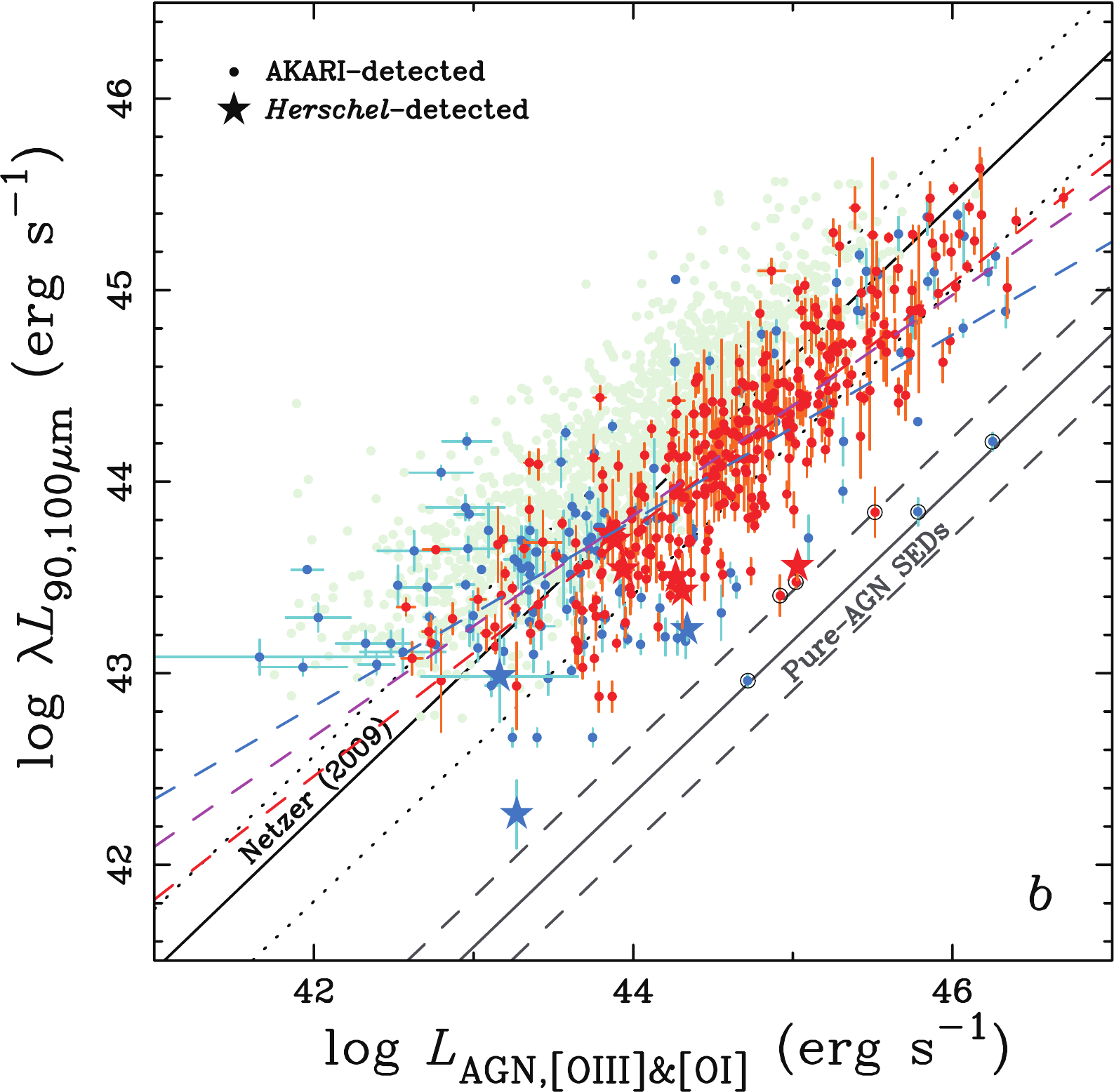}
\caption{Relations between FIR luminosity at 90 or 100 \micron, and AGN luminosity estimated from the [\ion{O}{3}]$\lambda$5007 line (the left-hand panel~(a)) and the combination of the [\ion{O}{3}]$\lambda$5007 and [\ion{O}{1}]$\lambda$6300 lines (the right-hand panel~(b)). The AKARI-detected and {\it Herschel}-detected objects are denoted with small circles and stars, respectively. Vertical and horizontal bars show 1$\sigma$ errors in their luminosities. Red and blue symbols indicate Seyfert 2s and LINER 2s, respectively. Blue-, red-, and purple-dashed lines are respective fitting results of Seyfert 2s, LINER 2s, and total objects. The reference line from \citet{2009MNRAS.399.1907N} is represented by black lines, assuming three different flux ratios (i.e., $F_{60\mu m}/F_{100\mu m}$), namely mean (the solid line), minimum and maximum ratios (dotted lines) from \citet{2002ApJ...576..159D}. The pure-AGN sequence with the 1$\sigma$ range is calculated from an intrinsic-AGN SED, shown as gray lines. Six pure-AGN candidates are denoted with large-black circles in the panel~(b). Light-green circles indicate composite objects selected with the BPT diagram, which are detected with AKARI.}
\label{bell}
\end{figure*}

Second, we compared FIR-based and [\ion{O}{2}]$\lambda$3727-based SFRs in Figure~\ref{fred}(b). SFRs are measured and calibrated based on H$\alpha$ and [\ion{O}{2}]$\lambda$3727 lines, mainly for normal galaxies \citep[e.g.,][]{1983ApJ...272...54K,1992ApJ...388..310K,1994ApJ...435...22K,1998ApJ...498..106M,2003ApJ...599..971H,2004AJ....127.2002K,2006ApJ...642..775M}. Recently, to derive attenuation-corrected line luminosities of galaxies, some studies have combined optical and infrared observations \citep[e.g.,][and references therein]{2009ApJ...703.1672K,2012MNRAS.426..330D}. Unfortunately, because of AGN contributions, it is difficult to adopt such corrections for our AGN sample. 
The best-fit relation between SFR$_{\rm FIR}$ and SFR$_{\rm [O\!~II]}$ (black-solid line) is steeper than the one-to-one correspondence, plausibly due to the combination of the following effects: the AGN contribution to the [\ion{O}{2}]$\lambda$3727 line, leading to an overestimation of the SFR, particularly for high [\ion{O}{2}]$\lambda$3727 luminosity objects, and the stronger aperture effect of the SDSS spectroscopy for lower-$z$, i.e., lower [\ion{O}{2}]$\lambda$3727 luminosity objects. Furthermore, for highly reddened objects, [\ion{O}{2}]$\lambda$3727-based SFR may become too high if extinction is over-corrected. It would be necessary to consider also metallicity and ionization conditions for the [\ion{O}{2}]$\lambda$3727 line. On the other hand, the FIR luminosity is contributed by dust in cirrus clouds that are warmed by diffuse starlight, particularly at low SFR (i.e., below 10 $M_\odot$/yr). This overestimates of SFR$_{\rm FIR}$ results in a steeper slope both for AGNs and normal star-forming galaxies.

Third, the UV-based and FIR-based SFRs are compared in Figure~\ref{fred}(c). Many objects are located below the one-to-one line, indicating that UV-based SFR is largely underestimated presumably due to the dust extinction. Note that since UV detection is not available for all AGNs, we include upper limits of UV-based SFR (open symbols) while most obscured galaxies are marked with green circles. The large scatter between UV-based and FIR-based SFRs (see Table~1), suggests that the UV-based SF are highly uncertain at any luminosity range, due to extinction and the contamination from AGB stars.

Fourth, Figure~\ref{fred}(d) presents a comparison of the UV-based SFR with SFR$_{\rm D4000}$, illustrating no significant correlation between them. On the other hand, Figure~\ref{fred}(e) shows a good correlation between the SFR$_{\rm UV}$ and SFR$_{\rm [O\!~II]}$. However, both SFRs suffer an extinction effect although there is a correlation between them.

Fifth, Figure~\ref{fred}(f) compares SFR$_{\rm D4000}$ and SFR$_{\rm [O\!~II]}$, showing a correlation with a systematic shift toward high SFR$_{\rm [O\!~II]}$. The larger SFR$_{\rm [O\!~II]}$ than SFR$_{\rm D4000}$ is mainly due to the AGN contribution to the [\ion{O}{2}]$\lambda$3727 line, while normal star-forming spirals show no excess of SFR$_{\rm [O\!~II]}$.

For quantitative analysis, we calculated the Spearman rank-order correlation coefficient $\rho$ and their statistical significance $p$ for all available data excluding upper limits. As presented in Table~1, we confirmed that the UV-based SFR seems to show week or almost no correlation with other indicators. FIR-based SFR seems to present slightly stronger correlations with optical-based SFRs than the relation among other SFRs, suggesting that FIR-based SFR is a reasonable SF indicator.

By comparing four independently estimated SFRs, we conclude that the FIR luminosity is the most reasonable SF indicator for host galaxies of type-2 AGNs. Thus, we will use the FIR luminosity as a SFR indicator, and compare it with the AGN luminosity in the following sections.

\subsection{A Relation between FIR and AGN Luminosities}\label{arel}
To examine the $L_{\rm FIR}$-$L_{\rm AGN}$ relation, we need to estimate AGN luminosities. X-ray luminosity is the most reliable indicator of AGN bolometric luminosity. However, it is not available for large samples including low-luminosity AGNs. The [\ion{Ne}{5}]$\lambda$3426 line is also a good indicator to estimate AGN luminosities without a contamination from \ion{H}{2} regions \citep[e.g.,][]{2010A&A...519A..92G}. Unfortunately, this emission line is typically weak and not covered by SDSS wavelength range for objects at $z < 0.1$. As a number of previous studies of type-2 AGNs used the [\ion{O}{3}]$\lambda$5007 line luminosity as a proxy for the AGN luminosities \citep[e.g.,][]{2004ApJ...613..109H,2006A&A...453..525N,2006MNRAS.372..961K,2009ApJ...695..793N,2009MNRAS.397..135K,2013ApJ...765L..33L}, we calculate bolometric luminosity from the [\ion{O}{3}]$\lambda$5007 line, adopting a bolometric correction, BC $= 600$ \citep[e.g.,][]{2009MNRAS.397..135K,2009MNRAS.399.1907N}. Figure~\ref{bell}(a) shows the relation between the FIR luminosity and AGN luminosity based on the [\ion{O}{3}]$\lambda$5007 line.

On the other hand, \citet{2009MNRAS.399.1907N} claimed that using the [\ion{O}{3}]$\lambda$5007 luminosity as a proxy for the AGN bolometric luminosity is unreliable due to its dependence on the ionisation parameter, which is critical for low-ionization sources such as LINERs. Thus, in order to avoid this ionization effect, we also calculated AGN bolometric luminosity from the combination of [\ion{O}{3}]$\lambda$5007 and [\ion{O}{1}]$\lambda$6300 line fluxes, using the calibration given by \citet{2009MNRAS.399.1907N}:
\begin{equation}
\log L_{\rm AGN} = 3.53 + 0.25 \log L_{\rm [O\!~III]} + 0.75 \log L_{\rm [O\!~I]},
\end{equation}
where $L_{\rm [O\!~III]}$ and $L_{\rm [O\!~I]}$ are extinction-corrected luminosities of [\ion{O}{3}]$\lambda$5007 and [\ion{O}{1}]$\lambda$6300 lines, respectively, in units of erg s$^{-1}$, using the Balmer decrement. Although the [\ion{O}{1}]$\lambda$6300 line is much weaker than [\ion{O}{3}]$\lambda$5007, we determined reliable AGN bolometric luminosities for 492 objects with S/N $> 3.0$ for the [\ion{O}{1}]$\lambda$6300 line. Note that even in the extreme case, e.g., S/N$_{\rm [O\!~III]} = 6$ and S/N$_{\rm [O\!~I]} = 3$, the uncertainty of logarithmic AGN luminosity, $\log L_{\rm AGN}$, is 0.25, and this is acceptable in our discussion. The relation between the FIR luminosity and AGN luminosity estimated by using the combination of two oxygen lines in Figure~\ref{bell}(b).

By comparing Figures~\ref{bell}(a) and \ref{bell}(b), we examine which AGN luminosity estimates is more reliable in this study. To investigate the ionization parameter effect, we plotted Seyfert 2s and LINER 2s with different symbols (i.e., red and blue symbols, respectively). We confirmed that the relation with the [\ion{O}{3}]$\lambda$5007-based AGN luminosity shows a larger scatter than that with AGN luminosity based on the combination of the [\ion{O}{3}]$\lambda$5007 and [\ion{O}{1}]$\lambda$6300 lines, particularly for LINER 2s at low $L_{\rm AGN}$ range. We performed a liner fit to Seyfert 2s, LINER 2s, and total objects, finding that the [\ion{O}{3}]$\lambda$5007 and [\ion{O}{1}]$\lambda$6300 combined method seems to correct for systematic trend in the distribution due to the ionization condition. By applying the Spearman rank-order test (see Table~1), we find a stronger correlation of FIR luminosity with AGN luminosity based on [\ion{O}{3}]$\lambda$5007 and [\ion{O}{1}]$\lambda$6300 method than based on [\ion{O}{3}]$\lambda$5007 only. These results imply that an ionization mechanism of Seyfert 2s and LINER 2s is similar. Thus, in the following analysis, we use both Seyfert 2s and LINER 2s and adopt the AGN bolometric luminosity estimated by [\ion{O}{3}]$\lambda$5007 and [\ion{O}{1}]$\lambda$6300 lines.

As shown in Figure~\ref{bell}, AKARI-detected objects show a clear trend between FIR and AGN luminosities, confirming the AGN-SF relation reported by previous studies. To directly compare with the previous studies, we included the reference line from \citet{2009MNRAS.399.1907N}, which represents the relation between the D4000-based SF luminosity and AGN luminosity. Note that the reference line was revised by converting 60 \micron\ luminosity to 90 \micron\ luminosity using three different flux ratios, i.e., $\log (F_{\rm 60\mu m}/F_{\rm 100\mu m}) = -0.55$, $-0.32$, and 0.21, which are based on the different SED templates \citep{2002ApJ...576..159D}. Our result is consistent with that of \citet{2009MNRAS.399.1907N} within the uncertainties of the FIR luminosity conversion. In contrast, we do not find a strong evidence of the enhanced SF for given AGN luminosity, as reported by \citet{2012A&A...545A..45R} for low-luminosity AGNs, particularly at high redshift, suggesting that low-luminosity AGNs are hosted by low-SF galaxies in the present day.

In addition, we plotted {\it Herschel}-detected objects as red stars. Since the flux limit of these objects are two orders of magnitude deeper than the AKARI/FIS survey, the additional {\it Herschel} sample helps us to overcome the flux limit of the shallow AKARI/FIS survey (see Figure~\ref{adam}). {\it Herschel}-detected objects are slightly shifted to lower $L_{\rm FIR}/L_{\rm AGN}$ ratio compared to AKARI-detected objects. Note that as we described in Section~\ref{sam} all type-2 AGNs in the COSMOS field are undetected with AKARI while most of them are detected with {\it Herschel}. On the other hand, $\sim$ 30,000 objects in the SDSS field are not detected with AKARI, implying that a large number of AGNs that are not detected with AKARI, may occupy the region where {\it Herschel}-detected objects are located in Figure~\ref{bell}.

\begin{figure}
\epsscale{1.0}
\plotone{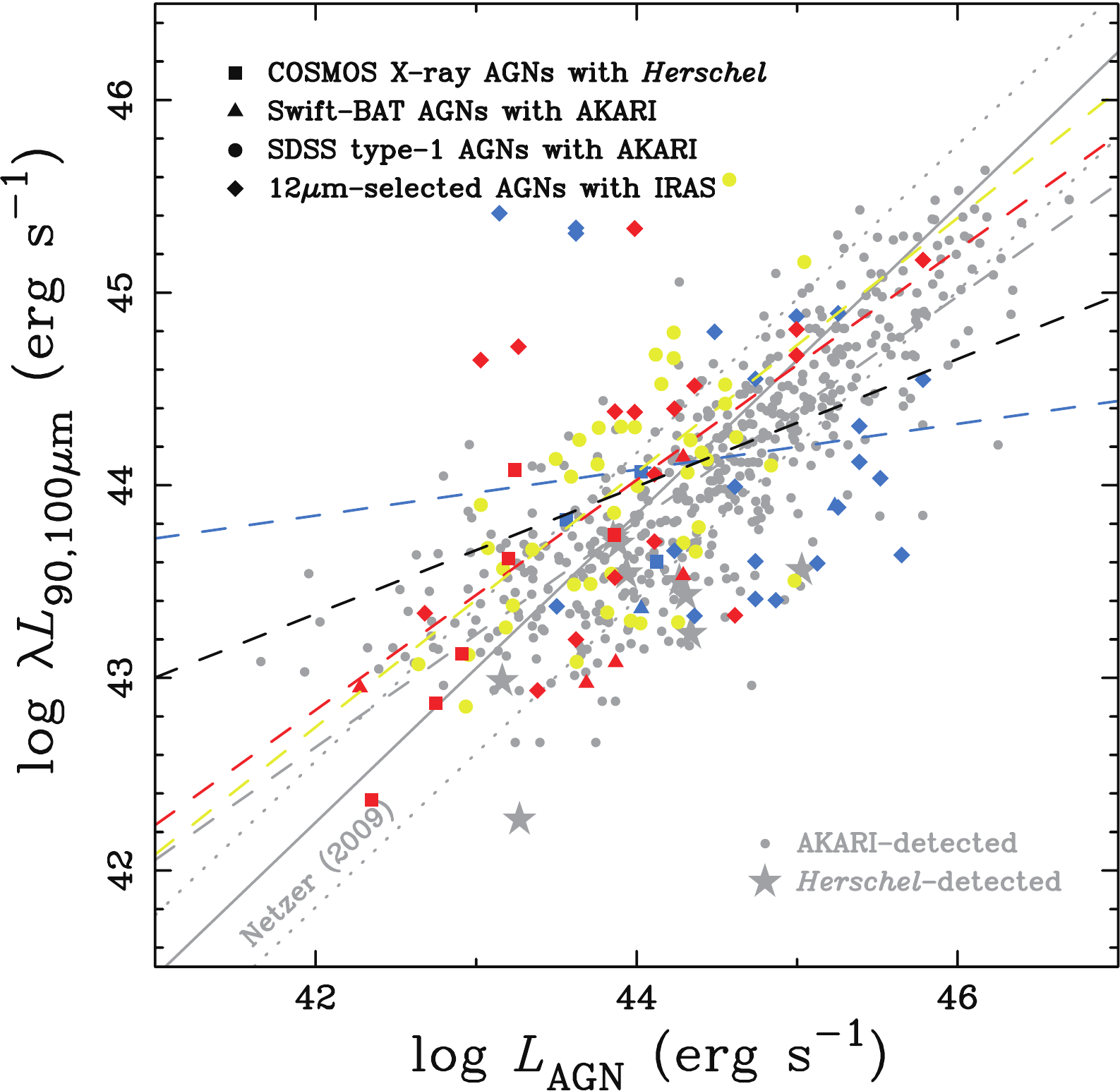}
\caption{Relation between FIR and AGN luminosities of type-1 and type-2 AGNs at $0.01 \le z < 0.22$. The gray circles, stars, and lines are same as in Figure~\ref{bell}(b). The X-ray type-2 and type-1 AGNs are represented by red and blue symbols, respectively. For each survey, individual symbols are given as labeled at the top left. In addition, SDSS type-1 AGNs with AKARI detections are also shown as yellow circles. Respective fitting results are shown as dashed lines with same colors of each symbol.}
\label{gray}
\end{figure}

\section{Discussion}\label{dis}
\subsection{Type-1 AGNs versus Type-2 AGNs}\label{anun}
We investigated the relation between FIR and AGN luminosities using type-2 AGNs, for which AGN bolometric luminosity is somewhat uncertain compared to type-1 AGNs. To overcome the uncertainty of the bolometric luminosity and to test whether type-1 AGNs also follow the same relation between FIR and AGN luminosities, we used X-ray AGN samples in this section.

First, we collected X-ray detected type-2 AGNs at $0.01 \leq z < 0.22$ from \citet{2011A&A...534A.110L}, and matched them against the PEP catalog, finding six {\it Herschel}-detected sources. For them, we adopted AGN bolometric luminosities estimated using infrared and X-ray luminosities \citep{2011A&A...534A.110L}. Second, we collected X-ray detected type-1 AGNs at the same redshift range from the COSMOS \citep{2010ApJ...716..348B}. By matching them against the PEP catalog, we obtained three {\it Herschel}-detected X-ray AGNs. Third, we collected X-ray detected AGNs from the 70 months Swift-BAT all-sky hard X-ray survey catalog \citep{2013MNRAS.431..836S}. By matching them against the AKARI/FIS catalog, we obtained two X-ray detected type-1 AGNs and four X-ray detected type-2 AGNs. For these AGNs, we estimated AGN bolometric luminosity from the X-ray luminosity with a bolometric correction in \citet{2009ApJ...700.1878R}. Fourth, we collected 23 Seyfert 1s and 18 Seyfert 2s from the 12 micron galaxy sample \citep{1989ApJ...342...83S,1993ApJS...89....1R}, which are detected in X-ray with {\it XMM-Newton} \citep{2011MNRAS.413.1206B}. Using these four samples, we plotted 28 type-1 AGNs (blue) and 28 type-2 AGNs (red) detected in X-rays on Figure~\ref{gray}. These X-ray AGNs seem to generally follow the similar trend between FIR and AGN luminosities, albeit the small sample size. To quantitatively assess these trends, we plotted the best-fit linear relations, respectively for X-ray type-2 AGNs and X-ray type-1 AGNs, and calculated the Spearman rank-order correlation coefficients (see Table~1). We find that X-ray type-2 AGNs show the similar relation to our SDSS type-2 AGNs, while X-ray type-1 AGNs seems to have a different slope. The shallow slope of X-ray type-1 AGNs seems due to the narrow $L_{\rm AGN}$ range. A larger sample is required to examine the origin of the difference in slope. Note that systematic errors from different data sets, especially in estimations of AGN luminosities, would affect these comparisons.

In addition to X-ray AGNs, we selected optically-selected type-1 AGNs using the SDSS low-luminosity AGN sample in \citet{2013MNRAS.431..836S}, for which the broad H$\alpha$ line is detected, enabling us to estimate AGN luminosity based on the H$\alpha$ line luminosity based on a recipe in \citet{2005ApJ...630..122G}. By matching them against the AKARI/FIS catalog, we obtained 45 type-1 AGNs. For these objects, an AGN bolometric luminosity is estimated from the broad H$\alpha$ line luminosity. In Figure~\ref{gray}, we plotted optical type-1 AGNs along with type-2 AGNs with yellow circles, showing that optical type-1 AGNs follow a consistent relation between FIR and AGN luminosities. We conclude the method of AGN bolometric luminosity estimation, i.e., narrow emission line luminosity, X-ray luminosity, and broad H$\alpha$ luminosity, does not significantly affect the relation between FIR and AGN luminosities, and that optical type-1 and type-2 AGNs show a similar relation between FIR and AGN luminosities.

\begin{figure}
\epsscale{1.0}
\plotone{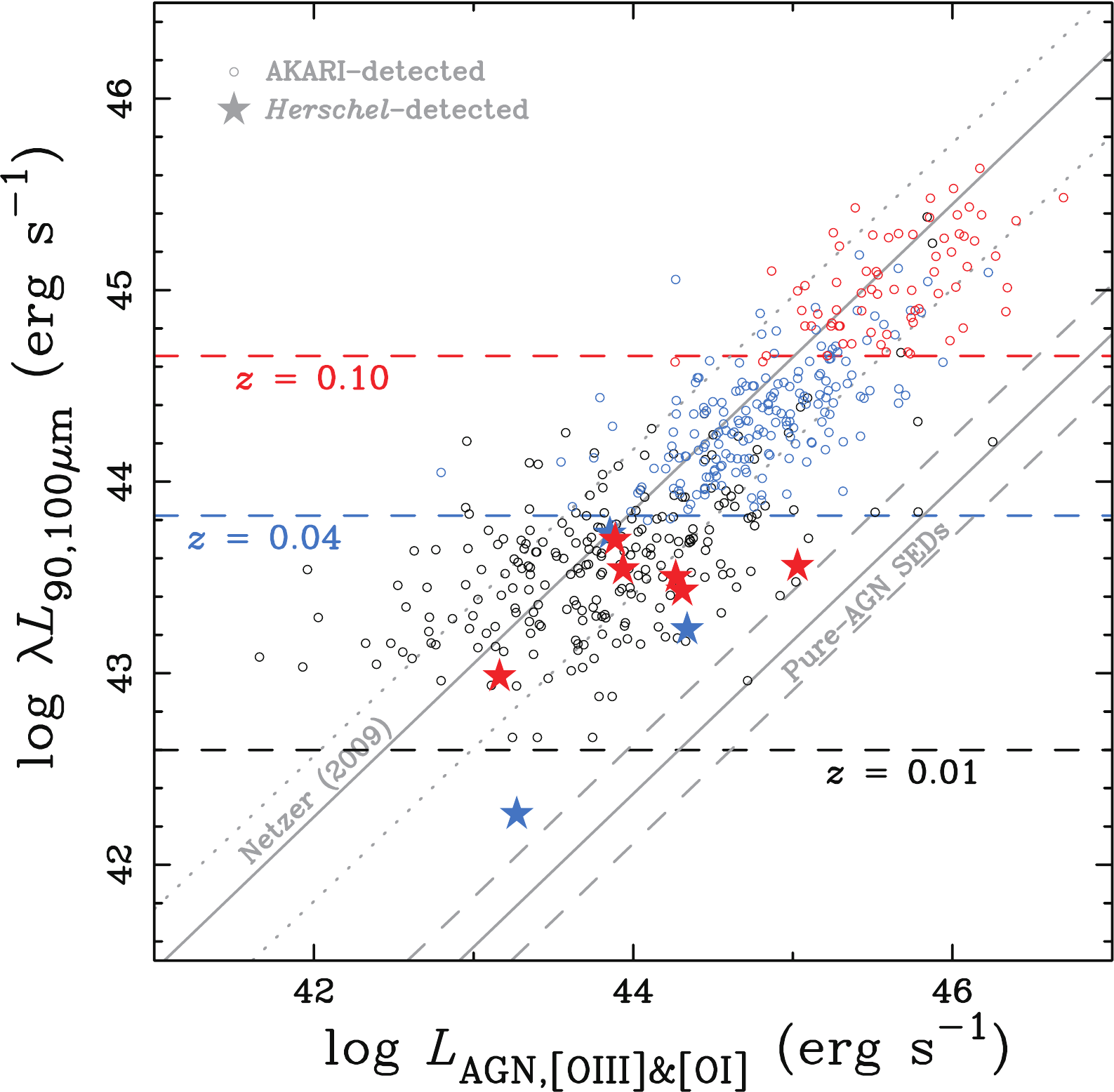}
\caption{Redshift distributions of our type-2 AGNs on the $L_{\rm FIR}$-$L_{\rm AGN}$ plane. The sample is divided in three redshift bins, $0.01 \leq z < 0.04$, $0.04 \leq z < 0.10$, and $0.10 \leq z < 0.22$, respectively denoted with black, blue, and red symbols. The AKARI-detected and {\it Herschel}-detected objects are shown as open circles and filled stars, respectively. The luminosity limits at $z =$ 0.01, 0.04, and 0.10 based on the AKARI 5$\sigma$-detection limits are denoted with horizontal dashed lines. Gray lines are the same as those in Figure~\ref{bell}.}
\label{dave}
\end{figure}

\begin{figure}
\epsscale{1.0}
\plotone{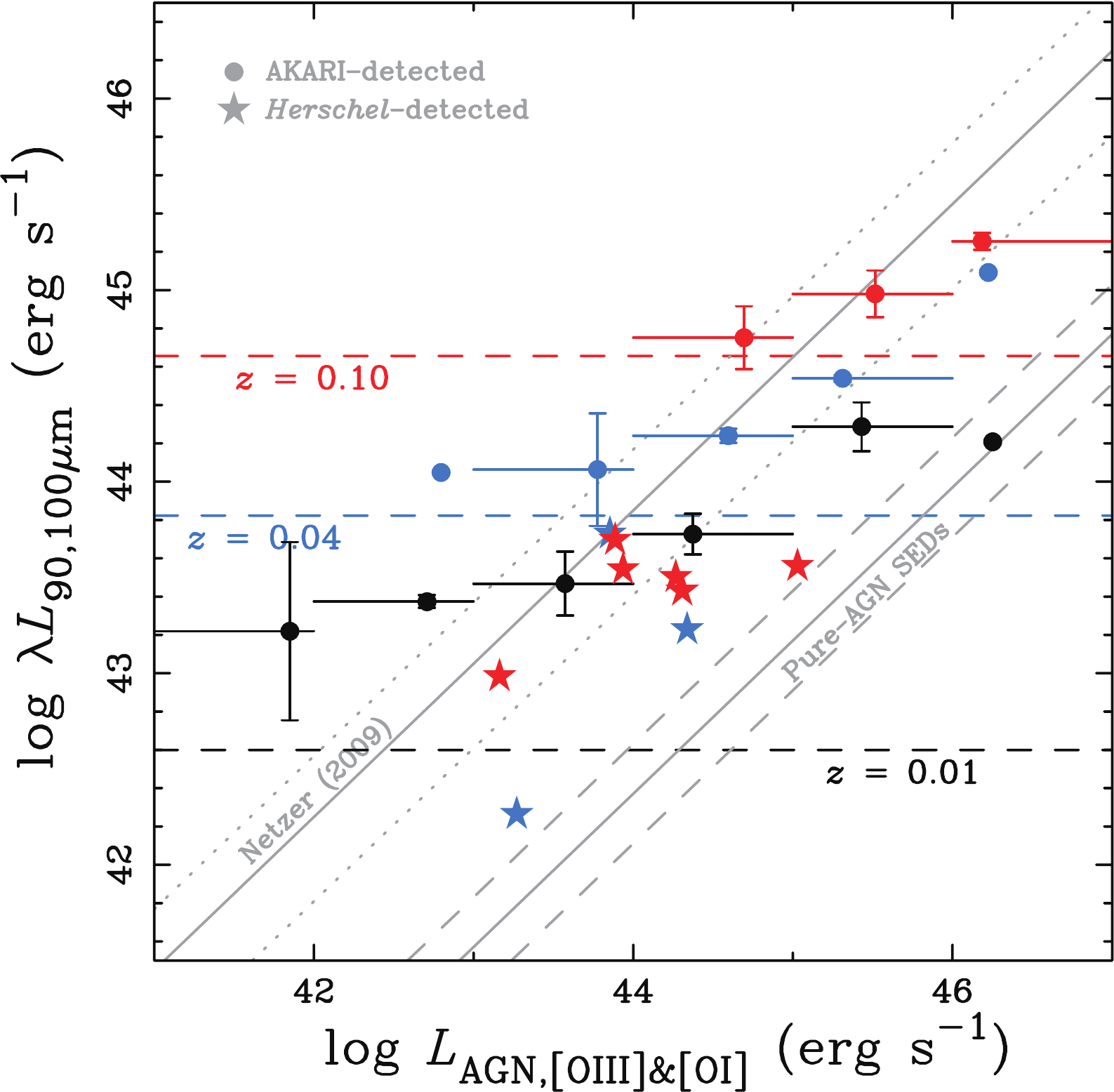}
\caption{Mean FIR luminosities of the AKARI-detected objects for each $L_{\rm AGN}$ bin. The mean luminosities in each redshift range, $0.01 \leq z < 0.04$, $0.04 \leq z < 0.10$, and $0.10 \leq z < 0.22$, are denoted with black, blue, and red circles, respectively. For each mean value, vertical bars represent 3$\sigma$ errors while horizontal bars show the AGN-luminosity ranges. If the sample size is less than 2 in a bin, a filled circle without error bars is given. The {\it Herschel}-detected objects are shown as filled stars. Horizontal and gray lines are the same as in Figure~\ref{dave}.}
\label{rose}
\end{figure}

\subsection{Comparisons with the Previous Studies}\label{comp}
In this section we compare our result with that of the previous studies. As shown in Figure~\ref{bell}, we found a similar relation between SF and AGN luminosities as reported by \citet{2009MNRAS.399.1907N}. The difference between our study and that of \citet{2009MNRAS.399.1907N} is that while we used FIR luminosity as a proxy for SF, \citet{2009MNRAS.399.1907N} used D4000 in estimating SF luminosity. As discussed in Section~\ref{star}, D4000-based SFR is not well determined at lower SFR since the calibration was based on starburst galaxies. Even with the FIR luminosity, which is a relatively better SF indicator, we find a similar relation between SF luminosity and AGN luminosity. This is probably due to the fact that the relation is not tight and the ratio between SF luminosity and AGN luminosity has a broad distribution, hence, the systematic difference between D4000-based SF luminosity and FIR luminosity is not clearly detected.

It is interesting to note that {\it Herschel}-detected objects follow the similar relation as AKARI-detected sources, i.e., low-$L_{\rm AGN}$ AGNs at higher redshift, show the similar trend between FIR and AGN luminosities, suggesting that the relation is not due to the selection effect. In Figure~\ref{dave}, the luminosity limits with increasing redshift is demonstrated in the $L_{\rm FIR}$-$L_{\rm AGN}$ plane. Here, we divided our sample into three redshift bins, i.e., $0.01 \leq z < 0.04$, $0.04 \leq z < 0.10$, and $0.10 \leq z < 0.22$, and the AKARI/FIS 5$\sigma$-detection limits at $z =$ 0.01, 0.04, and 0.10 are denoted with dashed horizontal lines. As shown in Figure~\ref{dave}, the AKARI-detected sources are strongly affected by the Malmquist bias, indicating that the AKARI/FIS sample alone does not allow us to investigate the relation between FIR and AGN luminosities without suffering the selection effect due to the flux limit. In contrast, {\it Herschel}-detected sources enable us to examine the relation over a wide luminosity range at given redshift (e.g., $0.1 \leq z < 0.22$ or $0.04 \leq z < 0.1$). In particular, for AGNs at $0.01 \leq z < 0.22$, the relation between FIR and AGN luminosities is detected over a wide range of AGN accretion luminosity, $42 \lesssim \log L_{\rm AGN}$ (erg s$^{-1}$) $\lesssim 46$. We conclude that the relation between FIR and AGN luminosities is not due to the flux limit of the AKARI/FIS surveys.

\begin{figure*}
\epsscale{1.0}
\plotone{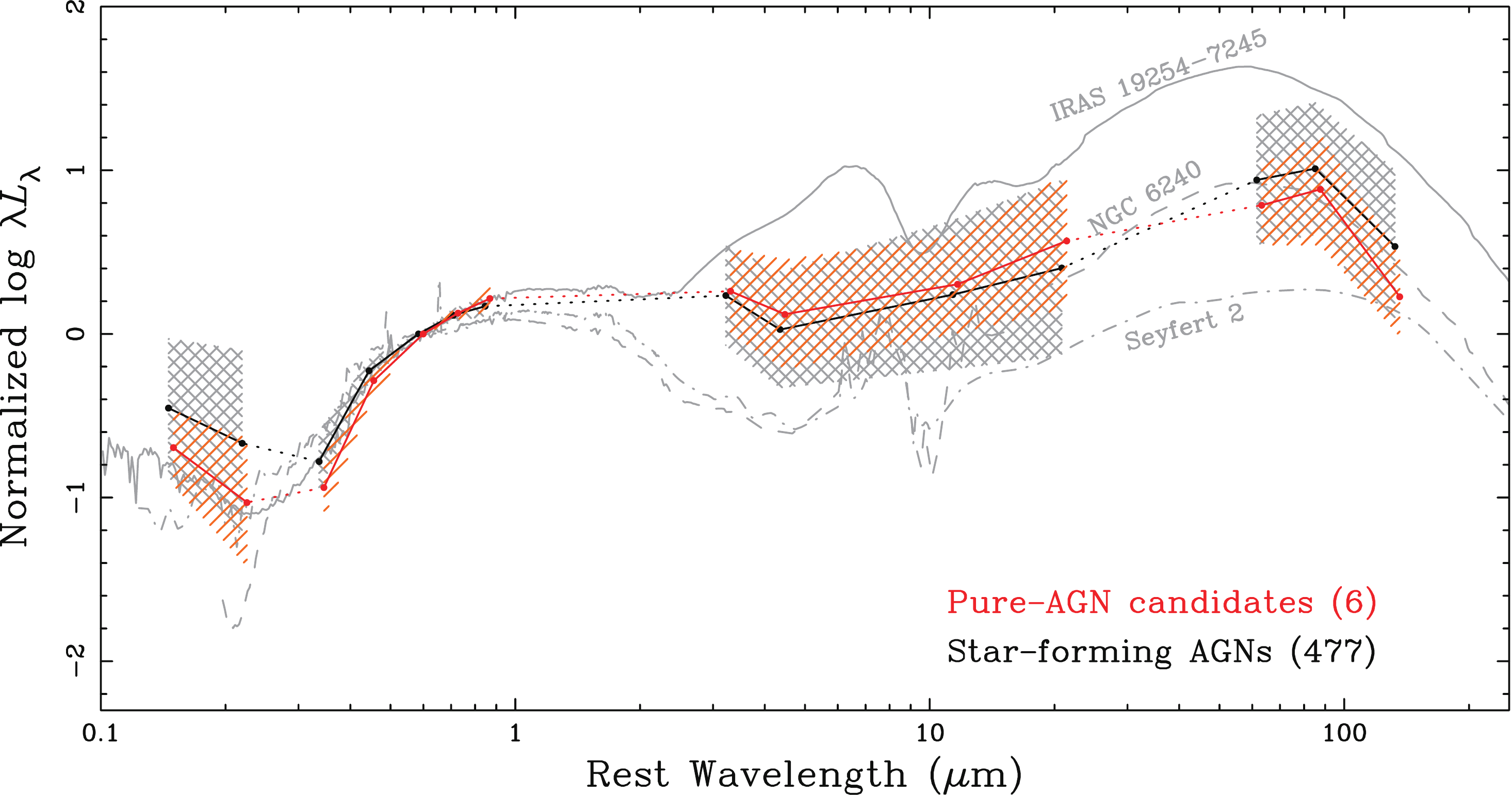}
\caption{Composite SEDs of pure-AGN candidates (red) and star-forming AGNs (black), normalized with SDSS $R$-band luminosity. Three representative SEDs are also shown as gray lines: starburst galaxy IRAS 19254$-$7245 (solid line), NGC 6240 (dashed line), and an average Seyfert 2 galaxy (dash-dot line).}
\label{dali}
\end{figure*}

To investigate the effect of the flux limit for given redshift bins, we calculated the mean FIR luminosities of the AKARI-detected sources in each redshift bin, i.e., $0.01 \leq z < 0.04$, $0.04 \leq z < 0.10$, and $0.10 \leq z < 0.22$, after dividing the AGNs in each redshift bin into subgroups based on AGN luminosity. In Figure~\ref{rose}, we present the mean FIR luminosities for each bin. The mean FIR and AGN luminosities show rather a flattened pattern in each redshift bin, compared to the relation between FIR and AGN luminosities of individual objects. In particular, at low AGN luminosity, the mean SF luminosity appears to be enhanced for fixed AGN luminosity, as similar reported by recent studies \citep[e.g.,][]{2010A&A...518L..26S,2012A&A...545A..45R}. However, this flattened pattern is not detected when we used individual luminosity measurements instead of mean luminosities \citep[see also][]{2009MNRAS.399.1907N}. The reason why we do not find AGNs hosted by galaxies with enhanced SF (i.e., above the one-to-one relation in Figure~\ref{bell}) may result from our sample selection since we excluded the composite objects in the BPT selection. It is possible that we missed SF-enhanced AGNs, which could be classified as composite objects, i.e., star-burst AGNs. To check whether composite objects are distributed above our relation or not, we plot galaxies classified as composite objects based on the BPT method (shown as light-green circles in Figure~\ref{bell}). Note that AGN bolometric luminosities might be overestimated for composite objects because of the contamination from star formation. As shown in Figure~\ref{gray}, we investigated the $L_{\rm AGN}$-$L_{\rm FIR}$ relation of X-ray selected objects and confirmed that several X-ray detected AGNs are distributed on the enhanced-SF area (top-left region in this figure), although its fraction is quite low, indicating that our results are consistent to \citet{2009MNRAS.399.1907N}. A direct test with the SDSS composite objects is difficult to perform since the AGN luminosity estimated from the narrow emission lines would be much more uncertain due to the contribution from SF. Note that \citet{2012A&A...545A..45R} adopted 60 \micron\ luminosity as a SF indicator which may be more contaminated by an AGN component than 90 and 100 \micron\ luminosities \citep[e.g.,][]{2002ApJ...572..105S}, and their mean FIR luminosities may be overestimated. Moreover, if \citet{2012A&A...545A..45R} lost low X-ray luminosity objects which would show low FIR luminosities in their sample selection, their averaged FIR luminosities would be overestimated although they calculated mean FIR luminosities by considering FIR detected and undetected objects.

\subsection{AGN Contribution to FIR Luminosities}\label{agnc}
If there are luminous AGNs hosted by low-SF galaxies, we may find them at the bottom right of the $L_{\rm FIR}$-$L_{\rm AGN}$ plane (see Figure~\ref{bell}). Usually, the AGN contribution to FIR is believed to be negligible, since SF galaxies dominates at FIR. When very low FIR luminosities are probed, however, it is necessary to quantify the AGN contribution. To investigate the FIR luminosities of a pure-AGN without SF, we adopted a SED template from \citet{2011MNRAS.414.1082M}. Using the strong correlation between the MIR (12.3 \micron) and the X-ray (2$-$10 keV) luminosities from \citet{2009A&A...502..457G}, we obtained the MIR luminosity as a function of AGN luminosity \citep[see also][]{2012ApJ...754...45I}, then calculate the FIR luminosity based on the pure-AGN SED template. In this process, we estimated the AGN bolometric luminosity from the X-ray luminosity \citep{2012A&A...545A..45R}.

In Figure~\ref{bell}, the estimated pure-AGN sequence is denoted with gray lines. Based on a comparison our sample with the pure-AGN sequence, we found that the AGN contribution in our sample seems to be negligible, although there are six objects reaching this pure-AGN sequence. Note that it is important to examine these objects located on the pure-AGN sequence since they are likely to be low-SF AGNs compared to the star-forming galaxies in our sample, if the expectation of contribution to FIR luminosities from the intrinsic-AGN SED is correct \citep[see also][]{2012A&A...545A..45R,2012MNRAS.420..526M}.

\begin{figure*}
\epsscale{0.49}
\plotone{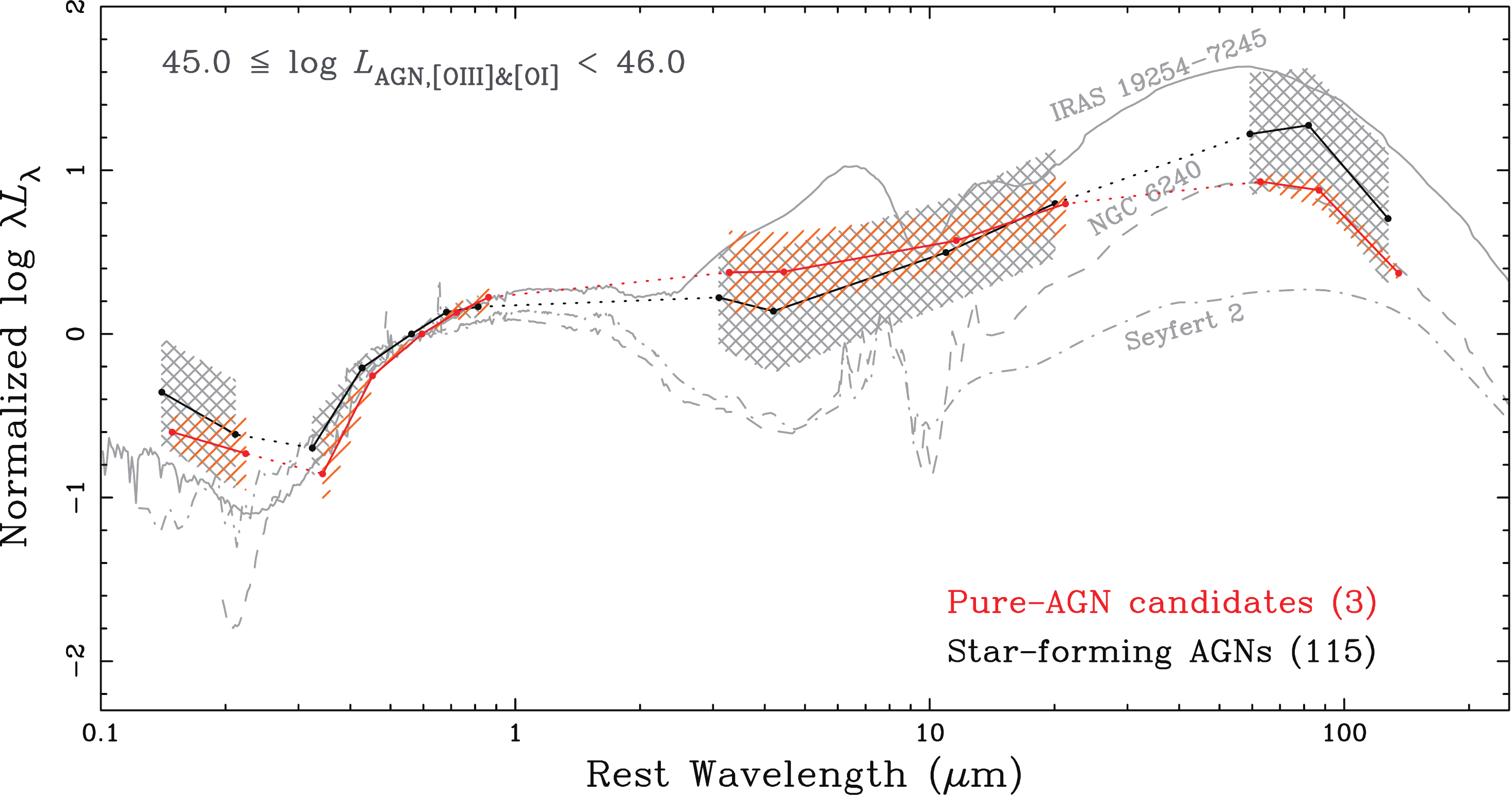}
\hspace{1.8mm}
\plotone{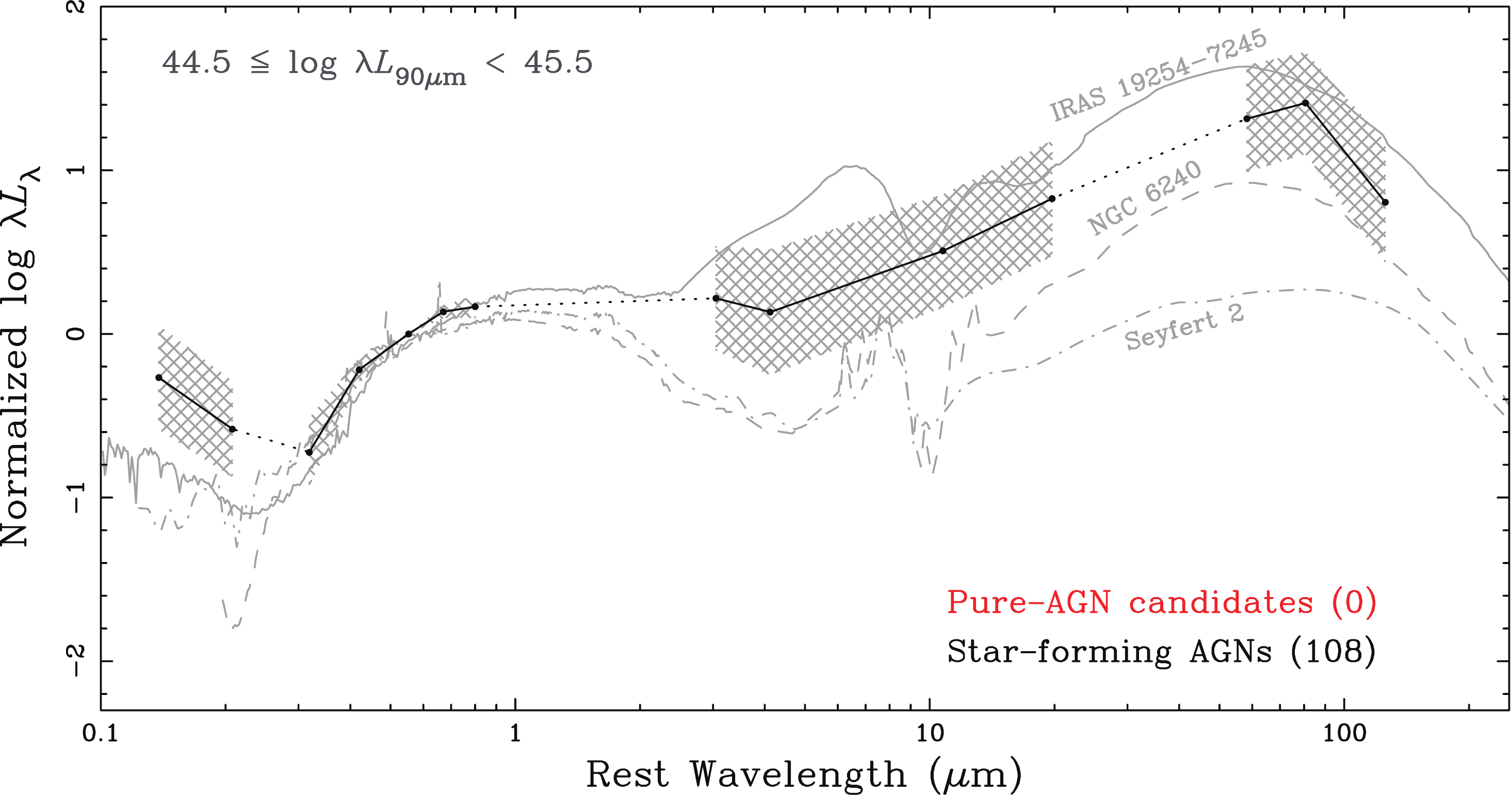}
\vspace{3.0mm}
\plotone{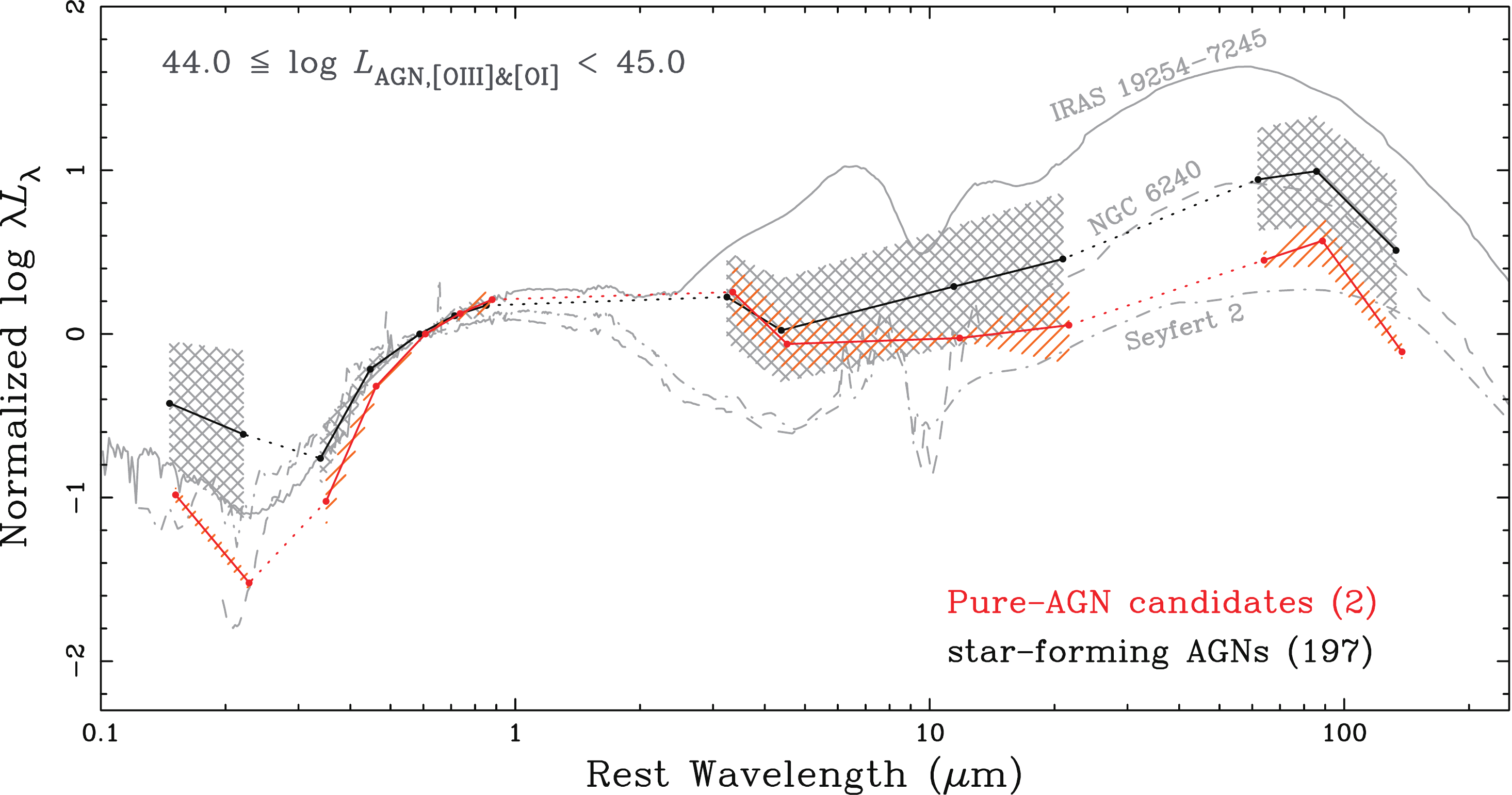}
\hspace{1.8mm}
\plotone{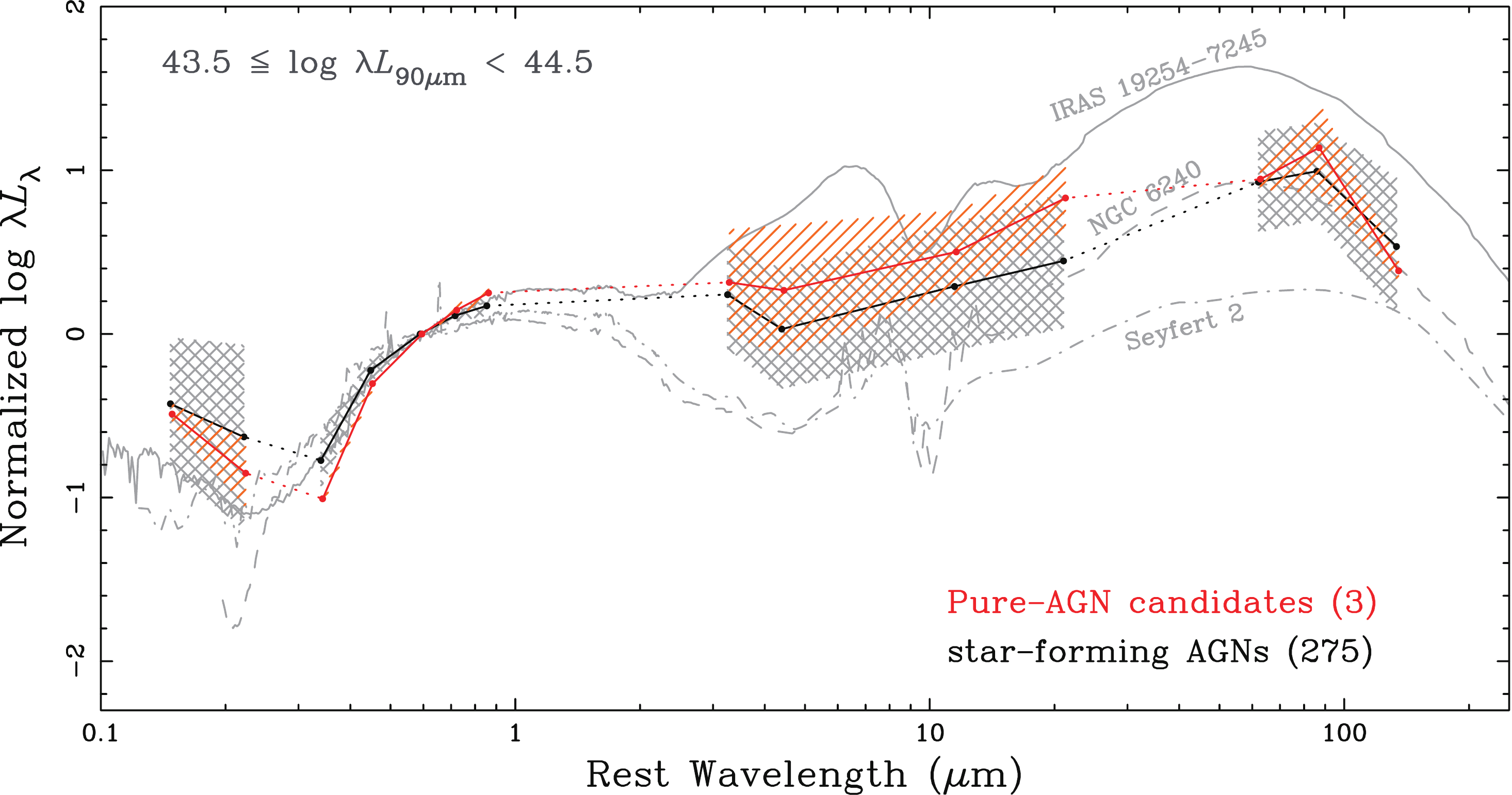}
\vspace{3.0mm}
\plotone{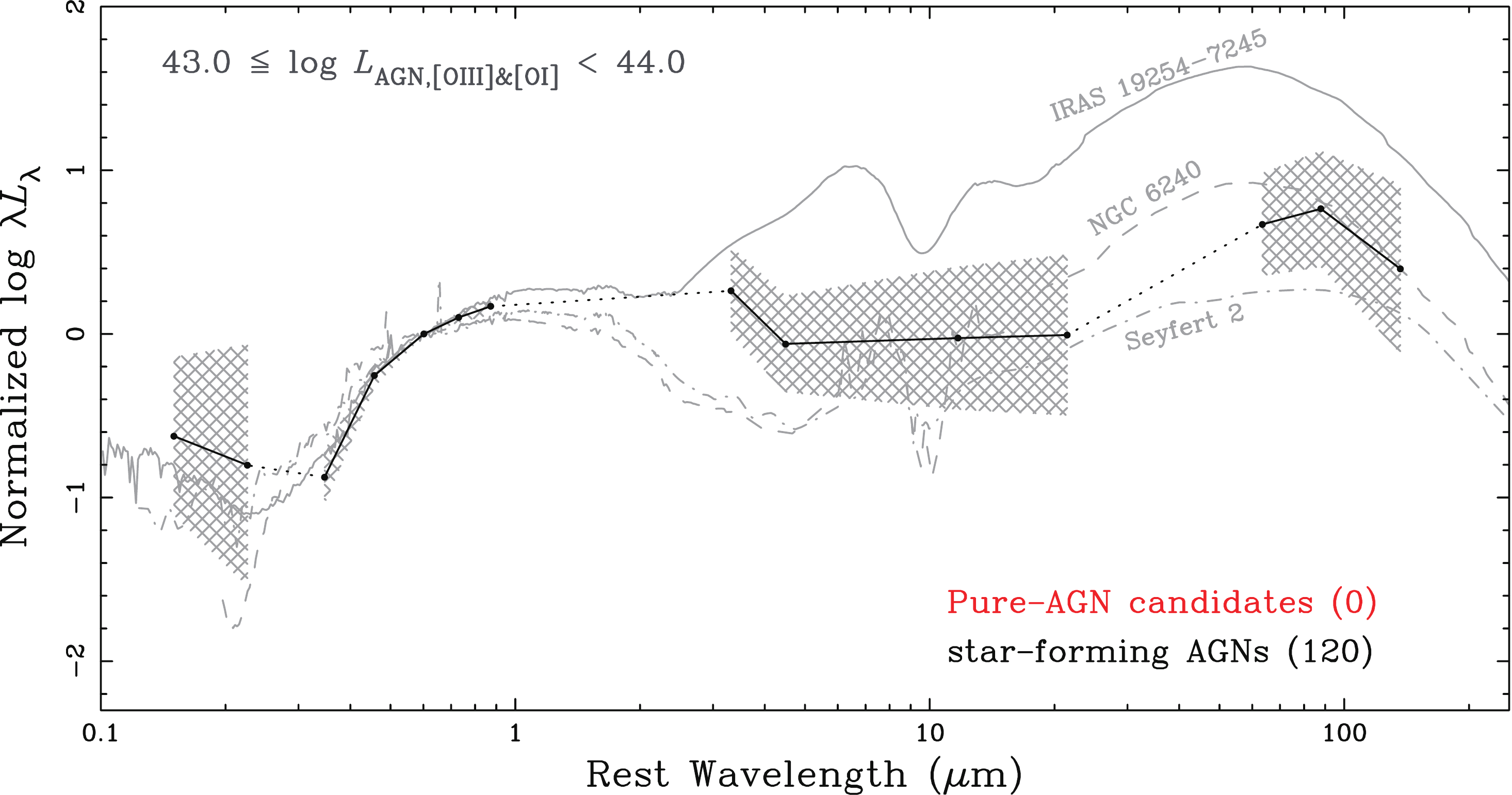}
\hspace{1.8mm}
\plotone{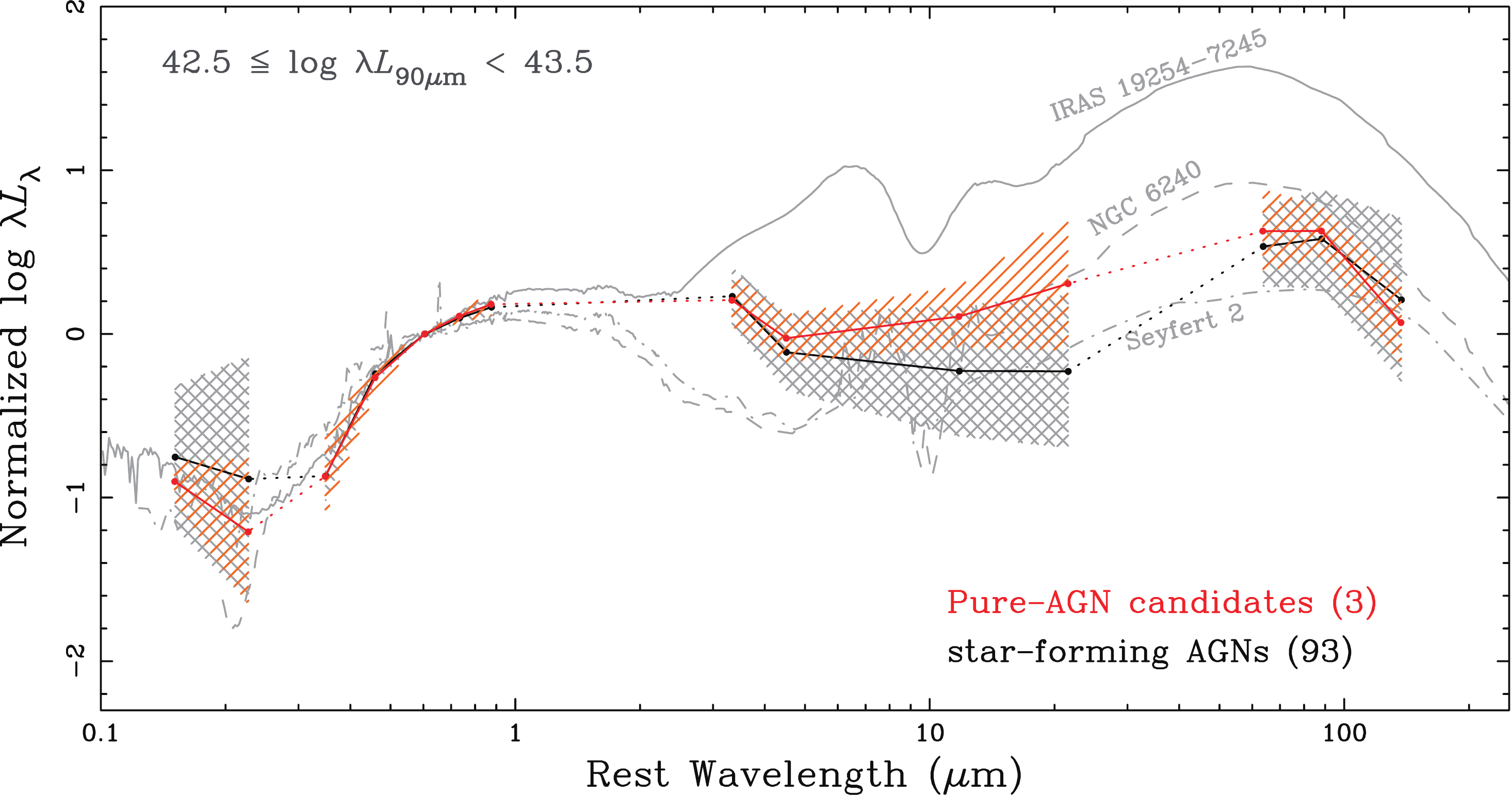}
\caption{Composite SEDs of pure-AGN candidates and star-forming AGNs, for three AGN-luminosity bins (left-hand panels), and for three FIR-luminosity bins (right-hand panels). The lines are the same as in Figure~\ref{dali}.}
\label{paul}
\end{figure*}

\subsection{Spectral Energy Distributions}\label{spec}
For understanding the $L_{\rm FIR}$-$L_{\rm AGN}$ relation in detail, in this section we investigate SEDs of our type-2 AGNs. As we mentioned in Section~\ref{agnc}, six objects are located close to the pure-AGN sequence. Here, we focus on these objects as pure-AGN candidates, i.e., luminous AGNs hosted by no- or low-SF galaxies (see large circles in Figure~\ref{bell}). Using multi-wavelength data, i.e., SDSS-optical, AKARI-FIR, {\it WISE}-MIR, and {\it GALEX}-UV data, we separately constructed the composite SEDs of pure-AGN candidates and AGNs hosted by star-forming galaxies shown in Figure~\ref{dali}. First, we normalized all-band luminosities with the SDSS $R$-band luminosity. Note that this normalization is effectively a stellar mass normalization, since optical-NIR luminosities roughly represent the stellar mass. In this construction, we used 5$\sigma$ upper limits for undetected sources. Along with the SEDs, we plotted three SED templates: a starburst galaxy IRAS 19254$-$7245, NGC 6240 (starburst + Seyfert 2), and a composite SED of Seyfert 2 galaxies \citep{2007ApJ...663...81P} as gray lines in Figure~\ref{dali}. These template SEDs are also normalized at the SDSS $R$-band wavelength.

As shown in Figure~\ref{dali}, we confirmed that the FIR luminosities of pure-AGN candidates are lower than AGNs hosted by star-forming galaxies, as already suggested in Figure~\ref{bell}. This indicates that pure-AGN candidates have significantly lower-SF host galaxies relative to star-forming AGNs. As SF indicators, UV luminosities also show a similar trend to the FIR luminosities, i.e., UV luminosities of pure-AGN candidates are lower than star-forming AGNs, although UV luminosities suffer from uncertainties as described in Section~\ref{star}. Moreover, in the optical range, there is a difference of the spectral slope between pure-AGN candidates and star-forming AGNs, suggesting the difference of D4000 between two groups. On the other hand, MIR luminosities, which are an indicator of the AGN luminosity, show no significant difference between pure-AGN candidates and star-forming AGNs. These results indicate that for given stellar mass, the AGN luminosity is comparable between pure-AGN candidates and star-forming AGNs while pure-AGN candidates have on average lower SF than star-forming AGNs. Compared to the templates, we found the SED of pure-AGN candidates is located between NGC 6240, which is a composite of starburst and Seyfert 2, and Seyfert 2 templates while the SED of star-forming AGNs is similar to the template of NGC 6240. This indicates that pure-AGN candidates are hosted by significantly lower-SF galaxies than star-forming AGNs. Note that even pure-AGN candidates have higher FIR luminosity than Seyfert 2 template, probably due to the AKARI flux-limit. In other words, AKARI sample is biased toward active star-forming hosts.

To understand the dependency of the SEDs in relation with $L_{\rm FIR}$ and $L_{\rm AGN}$, we divided our sample into subsamples by using FIR luminosities or accretion luminosities: we adopted AGN-luminosity bins of $43.0 \leq \log L_{\rm AGN} < 44.0$, $44.0 \leq \log L_{\rm AGN} < 45.0$, and $45.0 \leq \log L_{\rm AGN} < 46.0$, and for FIR-luminosity bins we used $42.5 \leq \log L_{\rm FIR} < 43.5$, $43.5 \leq \log L_{\rm FIR} < 44.5$, and $44.5 \leq \log L_{\rm FIR} < 45.5$. For each bin, we constructed an average SED
as plotted in Figure~\ref{paul}.

In the left hand panels, the SEDs for three $L_{\rm AGN}$ bins are shown from bottom to top with increasing accretion luminosities. In the case of star-forming AGNs shown as black lines, FIR and UV luminosities are increasing with increasing AGN luminosities. as expected from the positive trend between FIR and AGN luminosities. Also, we found that more luminous objects show slightly flatter slopes in the optical range, again suggesting the correlation between SF and AGN luminosities at fixed stellar mass. Moreover, we confirmed that the {\it WISE} continuum seems to increase with AGN luminosities, indicating that MIR bands would also be good $L_{\rm AGN}$ indicators \citep[e.g.,][]{2009A&A...502..457G}. For each AGN luminosity bin (top and middle panels) the SEDs of pure-AGN candidates show lower-SF activities than star-forming AGNs, which is consistent with the result in Figure~\ref{dali}.

\begin{figure*}
\epsscale{0.49}
\plotone{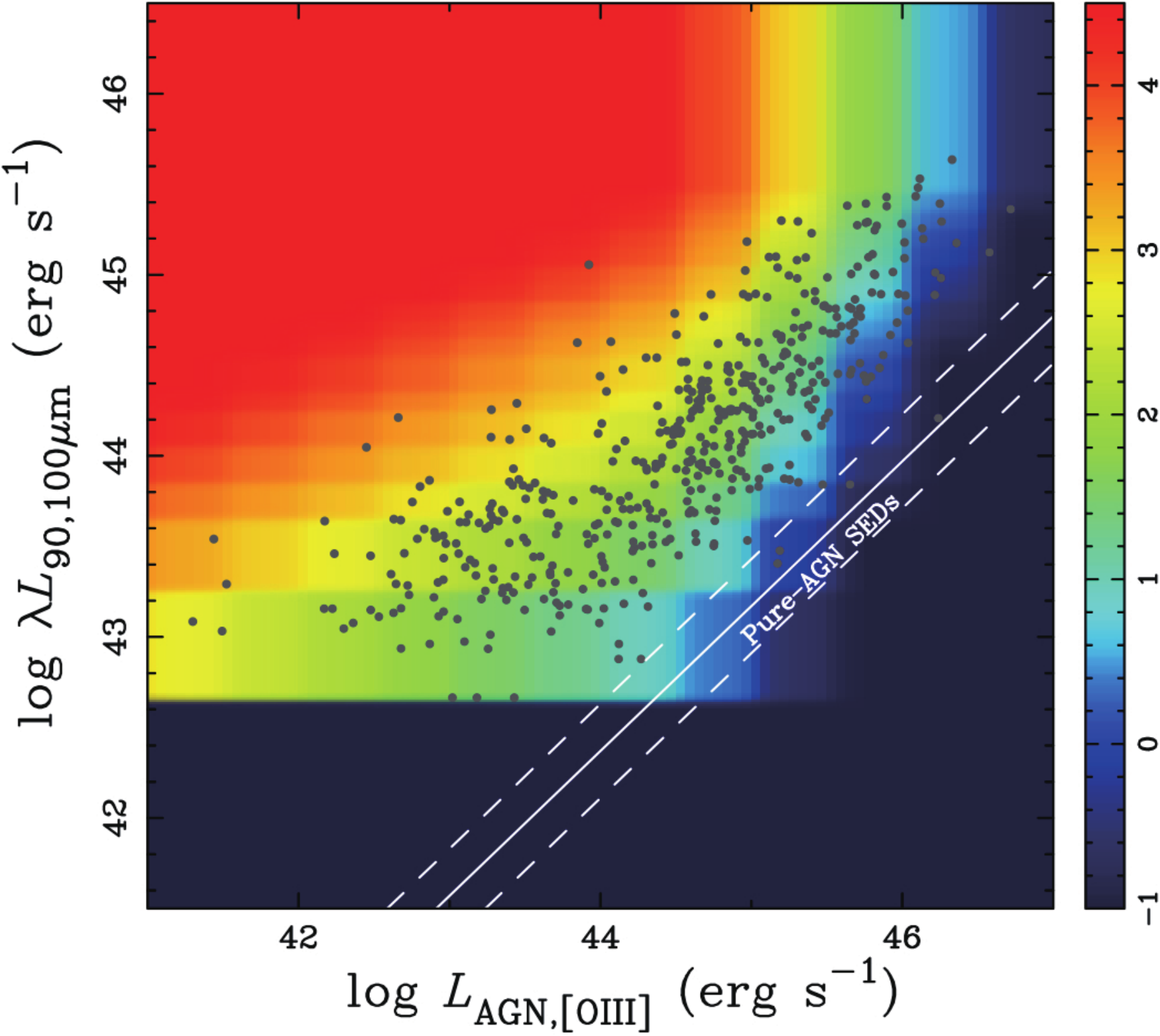}
\hspace{0.5mm}
\plotone{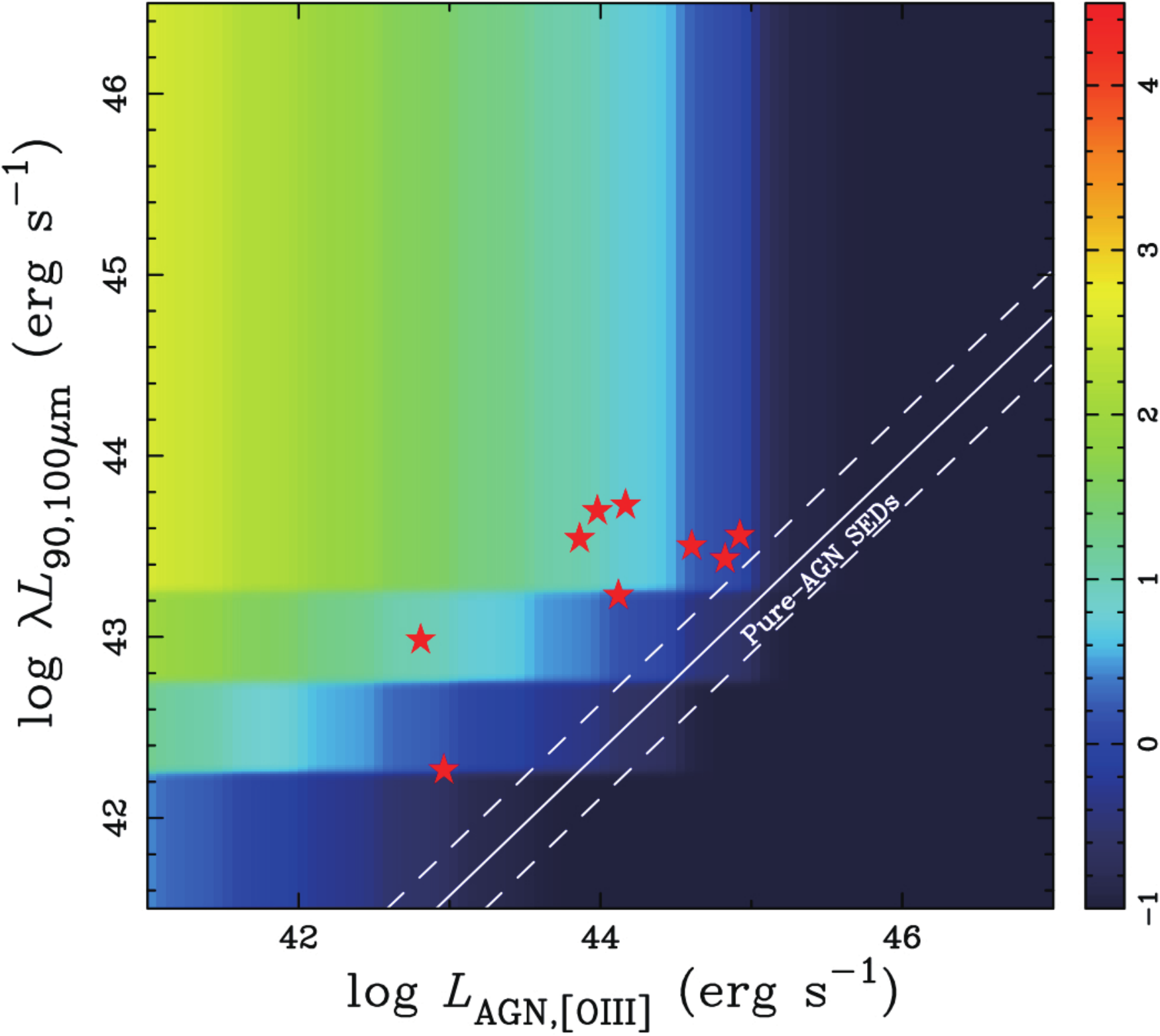}
\caption{Simulations of the $L_{\rm FIR}$-$L_{\rm AGN}$ distribution, demonstrating the effect of the flux limits in the AKARI/FIS all sky survey (left-hand panel), and the limited volume of the {\it Herschel}/PACS survey (right-hand panel). The logarithmic number density is calculated for each area bin with a size $\Delta \log L_{\rm FIR} = 0.2$ and $\Delta \log L_{\rm AGN} = 0.1$, and represented with different colors. For comparison, the AKARI-detected and {\it Herschel}-detected sources are also plotted with gray circles and red stars, respectively, while the pure-AGN sequence is denoted with white lines.}
\label{eddy}
\end{figure*}

When we divided the sample into three $L_{\rm FIR}$ bins as plotted in the right hand panels, we found that FIR and UV luminosities (i.e., SF indicators), and MIR luminosities (i.e., a tracer of AGN accretion) are increasing together as expected from the positive $L_{\rm FIR}$-$L_{\rm AGN}$ trend. The spectral slopes in the optical range are increasing with increasing FIR luminosities, reflecting the decrease of the D4000 (increase of SF). Note that the spectral slopes in the MIR range become steeper with increasing FIR luminosities, implying that there is a SF contribution to MIR luminosities. In the case of pure-AGN candidates (middle and bottom panels), pure-AGN candidates show higher MIR luminosities than star-forming AGNs, as expected from the trend in Figure~\ref{bell} that pure-AGN candidates have higher AGN luminosities than star-forming AGNs at fixed FIR luminosities.

Based on the SED analysis with FIR and AGN luminosity bins, we confirmed that pure-AGN candidates are hosted by low-SF galaxies compared to star-forming AGNs. Thus, these pure-AGN candidates could be a crucial sample for understanding the AGN-SF connection. Since the fraction of such objects appears to be small, $\sim 1\%$, this population may not be dominant in black hole growth history. However, it is possible that we are missing low-SF AGNs on the $L_{\rm FIR}$-$L_{\rm AGN}$ plane due to the observational limitations (see Section~\ref{arti}).

\subsection{The Effects of the Flux Limit and Volume Limit}\label{arti}
To better understand the observed relation between FIR and AGN luminosities in Figure~\ref{bell}, in this section we investigate the effects of the flux and volume limits by simulating the number density in the $L_{\rm FIR}$-$L_{\rm AGN}$ plane. Observationally it is difficult to find objects with high $L_{\rm AGN}$ and low $L_{\rm FIR}$ due to the following two effects:
\begin{itemize}
\item[--] at lower redshift, e.g., $z < 0.04$, it is difficult to detect high-$L_{\rm AGN}$ AGNs due to the limited survey volume,
\item[--] at higher redshift, e.g., $0.10 \leq z < 0.22$, objects with high $L_{\rm AGN}$ are easier to detect, however, the flux limits of the FIR survey prevent the detection of low-$L_{\rm FIR}$ galaxies hosting high-$L_{\rm AGN}$ AGNs.
\end{itemize}
To quantify these effects, we simulate the distribution of AGNs as a function of AGN luminosity using the [\ion{O}{3}]$\lambda$5007 luminosity function derived from the COSMOS and SDSS type-2 AGN samples \citep{2010A&A...510A..56B}. First, we calculated survey volumes for each redshift bin using the survey areas of the SDSS DR7 (i.e., 9380 deg$^2$) and the PEP-COSMOS field (i.e., 2.0069 deg$^2$), respectively for the AKARI and {\it Herschel} samples. Then, using the flux limits, i.e., 0.55 Jy for the AKARI/FIS survey and 7.5 mJy for the {\it Herschel}/PACS survey, we estimated object numbers for each area box with a fixed size ($\Delta \log L_{\rm FIR} = 0.2$ and $\Delta \log L_{\rm AGN} = 0.1$), with increasing redshift. Finally, we integrated the number of objects over the redshift range $0.01 \leq z < 0.22$.

Figure~\ref{eddy} presents the simulation results for the AKARI survey and the {\it Herschel} survey along with the observations. For simplicity, we adopted a [\ion{O}{3}]$\lambda$5007 line as a proxy for the AGN bolometric luminosity with a bolometric correction, BC $= 600$ \citep[e.g.,][]{2009MNRAS.397..135K,2009MNRAS.399.1907N}, since we used the [\ion{O}{3}]$\lambda$5007 luminosity function. The simulated distribution well reproduces the observed $L_{\rm FIR}$-$L_{\rm AGN}$ relation. Note that since we did not include the FIR luminosity function in our simulations, the top-left area shows very high object numbers. However, if we apply a FIR luminosity function, the object number in this area would decrease to zero.

Our simulation clearly indicates that the flux limit of the AKARI/FIS survey is insufficient to explore AGNs hosted by low-SF galaxies, close to the pure-AGN sequence while the limited volume of the PEP-COSMOS survey prevents us from detecting high $L_{\rm AGN}$ and low $L_{\rm FIR}$ sources. We conclude that the combination of the AKARI-detected and {\it Herschel}-detected AGNs used for our investigation suffers the observational limitations due to the flux limit and the survey volume. Thus, we were not able to investigate the number density of the pure AGNs or AGNs in post-starburst galaxies with the AKARI/FIS and {\it Herschel}/PACS data. For better understanding the $L_{\rm FIR}$-$L_{\rm AGN}$ relation, it is required to have a wide and deep FIR surveys with the next-generation infrared astronomy missions, e.g., the Space Infrared Telescope and Cosmology and Astrophysics (SPICA).

\section{Summary and Conclusions}\label{con}
To understand the AGN-SF connection, we investigated the relation between AGN and SF luminosities for a sample of SDSS type-2 AGNs at $z < 0.22$, based on the AKARI/FIS all-sky survey and the PEP COSMOS survey. We estimated AGN luminosities from [\ion{O}{3}]$\lambda$5007 and [\ion{O}{1}]$\lambda$6300 emission lines, and utilized the proposed linear proportionality of SFR with FIR luminosities in Kennicutt's equation. The main results are summarized as follows.
\begin{enumerate}
\item By comparing four independent SF indicators, i.e., FIR-based, UV-based, D4000-based, and [\ion{O}{2}]$\lambda$3727-based SFRs, we find that the FIR luminosity is the most acceptible and less subjective to AGN contamination compared to other SF indicators (Section~\ref{star}).
\item There is an apparent positive trend between FIR and AGN luminosities for local type-2 AGNs. In contrast to other studies \citep{2012A&A...545A..45R}, we find that low-$L_{\rm AGN}$ AGNs also follow the similar relation between FIR and AGN luminosities (Section~\ref{arel}).
\item Using X-ray AGNs and optical type-1 AGNs, we find a similar relation between $L_{\rm FIR}$ and $L_{\rm AGN}$, suggesting that the observed relation is not significantly affected by the method and uncertainty of the AGN bolometric luminosity estimation (Section~\ref{anun}).
\item The flux limit of AKARI FIR survey significantly affects the distribution in the $L_{\rm FIR}$-$L_{\rm AGN}$ plane while the deep-FIR data such as {\it Herschel} survey can overcome the limitation. It is possible that the AKARI FIR-detection limit is responsible for the observed trend in $L_{\rm FIR}$ and $L_{\rm AGN}$ plane (Section~\ref{comp}).
\item FIR luminosities of most type-2 AGNs are dominated by non-AGN continuum, while the AGN contributeion to FIR emission is negligible (Section~\ref{agnc}).
\item Based on the simulation of the AGN number distribution, we showed that the observed $L_{\rm FIR}$-$L_{\rm AGN}$ relation can be explained by the flux limit of the AKARI/FIS survey and the limited volume of the {\it Herschel}/PACS survey, demonstrating the limitations of the current survey data for detecting and investigating luminous AGNs hosted by low-SF galaxies (Section~\ref{arti}).
\end{enumerate}
Although it is possible that the observational limitations may cause an artificial correlation between FIR and AGN luminosities, the observed positive relation may suggest an intrinsic connection between SF and AGN activities in the present-day, implying that the growth of stellar mass and black hole mass are linked at least in the AGN phase in galaxy evolution. Quantifying the number density of luminous AGNs in low-SF or non-SF galaxies requires wide and deep future FIR surveys, e.g., SPICA.

\acknowledgments
We would like to thank Tohru~Nagao and Hagai~Netzer for their helpful comments and suggestions, and thank Hyun-Jin~Bae for the contribution of the line flux measurements. We also thank Matthew~A.~Malkan for his useful comments and suggestions which improved the clarity our paper. We also thank David~Rosario for his useful comments.
This work has been supported by the National Research Foundation of Korea (NRF) grant funded by the Korea government (No. 2012-006087 and No. 2013-K2A1A2055130).
K.~M. acknowledges financial support from the Japan Society for the Promotion of Science (JSPS).
Data analysis were in part carried out on common-use data analysis computer system at the Astronomy Data Center, ADC, of the National Astronomical Observatory of Japan (NAOJ).
Funding for the SDSS and SDSS-II has been provided by the Alfred P. Sloan Foundation, the Participating Institutions, the National Science Foundation, the U.S. Department of Energy, the National Aeronautics and Space Administration, the Japanese Monbukagakusho, the Max Planck Society, and the Higher Education Funding Council for England. The SDSS Web Site is http://www.sdss.org/. The SDSS is managed by the Astrophysical Research Consortium for the Participating Institutions. The Participating Institutions are the American Museum of Natural History, Astrophysical Institute Potsdam, University of Basel, University of Cambridge, Case Western Reserve University, University of Chicago, Drexel University, Fermilab, the Institute for Advanced Study, the Japan Participation Group, Johns Hopkins University, the Joint Institute for Nuclear Astrophysics, the Kavli Institute for Particle Astrophysics and Cosmology, the Korean Scientist Group, the Chinese Academy of Sciences (LAMOST), Los Alamos National Laboratory, the Max-Planck-Institute for Astronomy (MPIA), the Max-Planck-Institute for Astrophysics (MPA), New Mexico State University, Ohio State University, University of Pittsburgh, University of Portsmouth, Princeton University, the United States Naval Observatory, and the University of Washington. 
This research is based on observations with AKARI, a JAXA project with the participation of ESA. 
{\it Herschel} is an ESA space observatory with science instruments provided by European-led Principal Investigator consortia and with important participation from NASA. 


\clearpage
\begin{turnpage}
\begin{deluxetable}{llcrrrcrrcrrclc}
\tabletypesize{\scriptsize}
\tablecaption{Results of Correlation Fittings and the Spearman Rank-order Test}
\tablewidth{0pc}
\tablehead{
 & & & \multicolumn{3}{c}{$\log y = a \log x + b$} & & \multicolumn{2}{c}{$\log y = \log x + c$}
 & & \multicolumn{2}{c}{Spearman} & & Sample & Figure\\
\noalign{\smallskip}\cline{4-6}\cline{8-9}\cline{11-12}\noalign{\smallskip}
\colhead{$x$} & \colhead{$y$} & & \colhead{$a$} & \colhead{$b$} & \colhead{rms} & & \colhead{$c$} & \colhead{rms}
 & & \colhead{$\rho$} & \colhead{$p$} & & &
}
\startdata
${\rm SFR}_{\rm D4000}$          & ${\rm SFR}_{\rm FIR}$   & &    1.085$\pm$0.041 & $-$0.464$\pm$0.039 & 0.569 & &
 $-$0.403$\pm$0.026 & 0.571 & & $+$0.79 & $< 2.20 \times 10^{-16}$ & & \multicolumn{1}{c}{--} & 2(a)\\
${\rm SFR}_{\rm [O\!~II]}$       & ${\rm SFR}_{\rm FIR}$   & &    1.281$\pm$0.054 &    0.075$\pm$0.058 & 0.546 & &
    0.334$\pm$0.030 & 0.565 & & $+$0.80 & $< 2.20 \times 10^{-16}$ & & \multicolumn{1}{c}{--} & 2(b)\\
${\rm SFR}_{\rm UV}$             & ${\rm SFR}_{\rm FIR}$   & &    0.621$\pm$0.137 &    0.454$\pm$0.137 & 1.387 & &
    0.164$\pm$0.089 & 1.406 & & $+$0.28 & $  1.06 \times 10^{-5}$ \hspace{0.8pt} & & \multicolumn{1}{c}{--} & 2(c)\\
${\rm SFR}_{\rm D4000}$          & ${\rm SFR}_{\rm UV}$    & &    0.091$\pm$0.035 &    0.443$\pm$0.059 & 0.786 & &
 $-$0.403$\pm$0.097 & 1.527 & & $+$0.18 & $  4.03 \times 10^{-3}$ \hspace{0.8pt} & & \multicolumn{1}{c}{--} & 2(d)\\
${\rm SFR}_{\rm [O\!~II]}$       & ${\rm SFR}_{\rm UV}$    & &    0.924$\pm$0.077 & $-$0.159$\pm$0.189 & 0.989 & &
 $-$0.014$\pm$0.069 & 0.989 & & $+$0.66 & $< 2.20 \times 10^{-16}$  & & \multicolumn{1}{c}{--} & 2(e)\\
${\rm SFR}_{\rm [O\!~II]}$       & ${\rm SFR}_{\rm D4000}$ & &    0.805$\pm$0.045 &    0.772$\pm$0.043 & 0.635 & &
    0.655$\pm$0.034 & 0.650 & & $+$0.73 & $< 2.20 \times 10^{-16}$ & & \multicolumn{1}{c}{--} & 2(f)\\
\noalign{\smallskip}\cline{1-15}\noalign{\smallskip}
$L_{\rm AGN,[O\!~III]}$          & $\lambda L_{\rm 90,100\mu m}$ & & 0.611$\pm$0.022 & 16.820$\pm$0.998 & 0.321 & &
 \multicolumn{1}{c}{--} & \multicolumn{1}{c}{--} & & $+$0.82 & $< 2.20 \times 10^{-16}$ & & Seyfert 2s & 3(a)\\
$L_{\rm AGN,[O\!~III]}$          & $\lambda L_{\rm 90,100\mu m}$ & & 0.464$\pm$0.038 & 23.509$\pm$1.650 & 0.470 & &
 \multicolumn{1}{c}{--} & \multicolumn{1}{c}{--} & & $+$0.61 & $< 2.20 \times 10^{-16}$ & & LINER 2s   & 3(a)\\
$L_{\rm AGN,[O\!~III]}$          & $\lambda L_{\rm 90,100\mu m}$ & & 0.494$\pm$0.017 & 22.099$\pm$0.776 & 0.384 & &
 \multicolumn{1}{c}{--} & \multicolumn{1}{c}{--} & & $+$0.81 & $< 2.20 \times 10^{-16}$ & & Total      & 3(a)\\
\noalign{\smallskip}\cline{1-15}\noalign{\smallskip}
$L_{\rm AGN,[O\!~III]\&[O\!~I]}$ & $\lambda L_{\rm 90,100\mu m}$ & & 0.644$\pm$0.022 & 15.414$\pm$0.961 & 0.302 & &
 \multicolumn{1}{c}{--} & \multicolumn{1}{c}{--} & & $+$0.85 & $< 2.20 \times 10^{-16}$ & & Seyfert 2s & 3(b)\\
$L_{\rm AGN,[O\!~III]\&[O\!~I]}$ & $\lambda L_{\rm 90,100\mu m}$ & & 0.485$\pm$0.037 & 22.457$\pm$1.642 & 0.456 & &
 \multicolumn{1}{c}{--} & \multicolumn{1}{c}{--} & & $+$0.63 & $< 2.20 \times 10^{-16}$ & & LINER 2s   & 3(b)\\
$L_{\rm AGN,[O\!~III]\&[O\!~I]}$ & $\lambda L_{\rm 90,100\mu m}$ & & 0.576$\pm$0.018 & 18.478$\pm$0.807 & 0.357 & &
 \multicolumn{1}{c}{--} & \multicolumn{1}{c}{--} & & $+$0.84 & $< 2.20 \times 10^{-16}$ & & Total      & 3(b)\\
\noalign{\smallskip}\cline{1-15}\noalign{\smallskip}
$L_{\rm AGN}$                    & $\lambda L_{\rm 90,100\mu m}$ & & 0.119$\pm$0.113 & 38.843$\pm$5.045 & 0.705 & &
 \multicolumn{1}{c}{--} & \multicolumn{1}{c}{--} & & $+$0.17 & $  3.79 \times 10^{-1}$ \hspace{0.8pt} & & X-ray AGN 1s & 4\\
$L_{\rm AGN}$                    & $\lambda L_{\rm 90,100\mu m}$ & & 0.597$\pm$0.171 & 17.746$\pm$7.498 & 0.679 & &
 \multicolumn{1}{c}{--} & \multicolumn{1}{c}{--} & & $+$0.54 & $  3.56 \times 10^{-3}$ \hspace{0.8pt} & & X-ray AGN 2s & 4\\
$L_{\rm AGN}$                    & $\lambda L_{\rm 90,100\mu m}$ & & 0.661$\pm$0.189 & 14.982$\pm$8.315 & 0.720 & &
 \multicolumn{1}{c}{--} & \multicolumn{1}{c}{--} & & $+$0.54 & $  1.42 \times 10^{-4}$ \hspace{0.8pt} & & SDSS AGN 1s  & 4\\
$L_{\rm AGN}$                    & $\lambda L_{\rm 90,100\mu m}$ & & 0.331$\pm$0.081 & 29.440$\pm$3.576 & 0.723 & &
 \multicolumn{1}{c}{--} & \multicolumn{1}{c}{--} & & $+$0.79 & $< 2.20 \times 10^{-16}$ & & Total & 4\\
\enddata
\end{deluxetable}
\end{turnpage}

\end{document}